\documentclass[referee]{aa}
\usepackage{psfig,amsmath}

\begin{document}

\thesaurus{05(03.20.4)} 

\title{Photometric study of the double cluster $h$ \& $\chi$ Persei\thanks{Based on observations obtained with the Jacobus Kapteyn Telescope operated on the island of La Palma by the Isaac Newton Group, in the Spanish Observatorio Roque de los Muchachos of the Instituto de Astrof\'{\i}sica de Canarias.}}

\author{Amparo Marco\inst{1}
\and Guillermo Bernabeu\inst{1}}
                                                            
\institute{Dpto. de F\'{\i}sica, Ingenier\'{\i}a de Sistemas y 
Teor\'{\i}a de la Se\~{n}al, Universidad de Alicante, Aptdo. de Correos 99,
E-03080, Alicante, Spain}

\mail{amparo@astronomia.disc.ua.es}

\date{Received 20 July 2000 / Accepted 28 January 2001}

\titlerunning{$uvby\beta$ CCD Photometry of $h$ \& $\chi$ Persei}
\authorrunning{Marco \& Bernabeu}
\maketitle 

\begin{abstract}
We present $uvby\beta$ CCD photometry of the central region of the 
double cluster $h$ \& $\chi$ Persei. We identify $\approx$ 350
stars, of which 214 were not included in Oosterhof's catalogue.
Our magnitude limit $V=16.5$ allows us to reach early F spectral type
and obtain very accurate fits to the ZAMS. We derive reddening values of
$E(b-y) = 0.44\pm0.02$ for $h$ Persei and $E(b-y) = 0.39\pm0.05$ for
$\chi$ Persei. From the ZAMS fitting, we derive distance moduli
$V_{0}-M_{V} = 11.66\pm0.20$ and $V_{0}-M_{V} = 11.56\pm0.20$ for
$h$ and $\chi$ Persei respectively. These values are perfectly compatible
with both clusters being placed at the same distance and having identical
reddenings. The shift in the main-sequence turnoff and isochrone 
fitting, however, show that there is a significant age difference between both
clusters, with the bulk of stars in $h$ Persei being older than $\chi$ 
Persei. There is, however, a significant population of stars in $h$ Persei
which are younger than $\chi$ Persei. All this argues for at least
three different epochs of star formation, corresponding approximately
to $\log t = 7.0, 7.15$ and $7.3$.
\end{abstract}

\keywords{techniques:photometry -- Galaxy:open clusters and 
associations:individual:$h$ Persei, $\chi$ Persei-- stars: evolution -- emission-line,
 Be--formation}

\section{Introduction}
Photometric studies of open clusters are extremely useful to 
determine the physical properties of star members. Even though modern
spectroscopic techniques allow the observation of large numbers of stars 
in a relatively short time, the stellar population of most clusters is still
too vast for an in-depth study. Among photometric systems, 
the $uvby$H$\beta$ system is the most appropriate for the study of
early-type stars, since it has been designed to provide accurate measurements
of their intrinsic properties.

This is the first in a series of papers dedicated to the study of the B-star 
population of Galactic open clusters. B-type stars are sufficiently bright to 
allow very accurate narrow-band photometry and at the same time numerous enough
to provide a statistically significant population (unlike the brighter but very
rare O-type star).
\begin{figure*}
\begin{center}
\begin{tabular}{cc}
&\\
\psfig{file=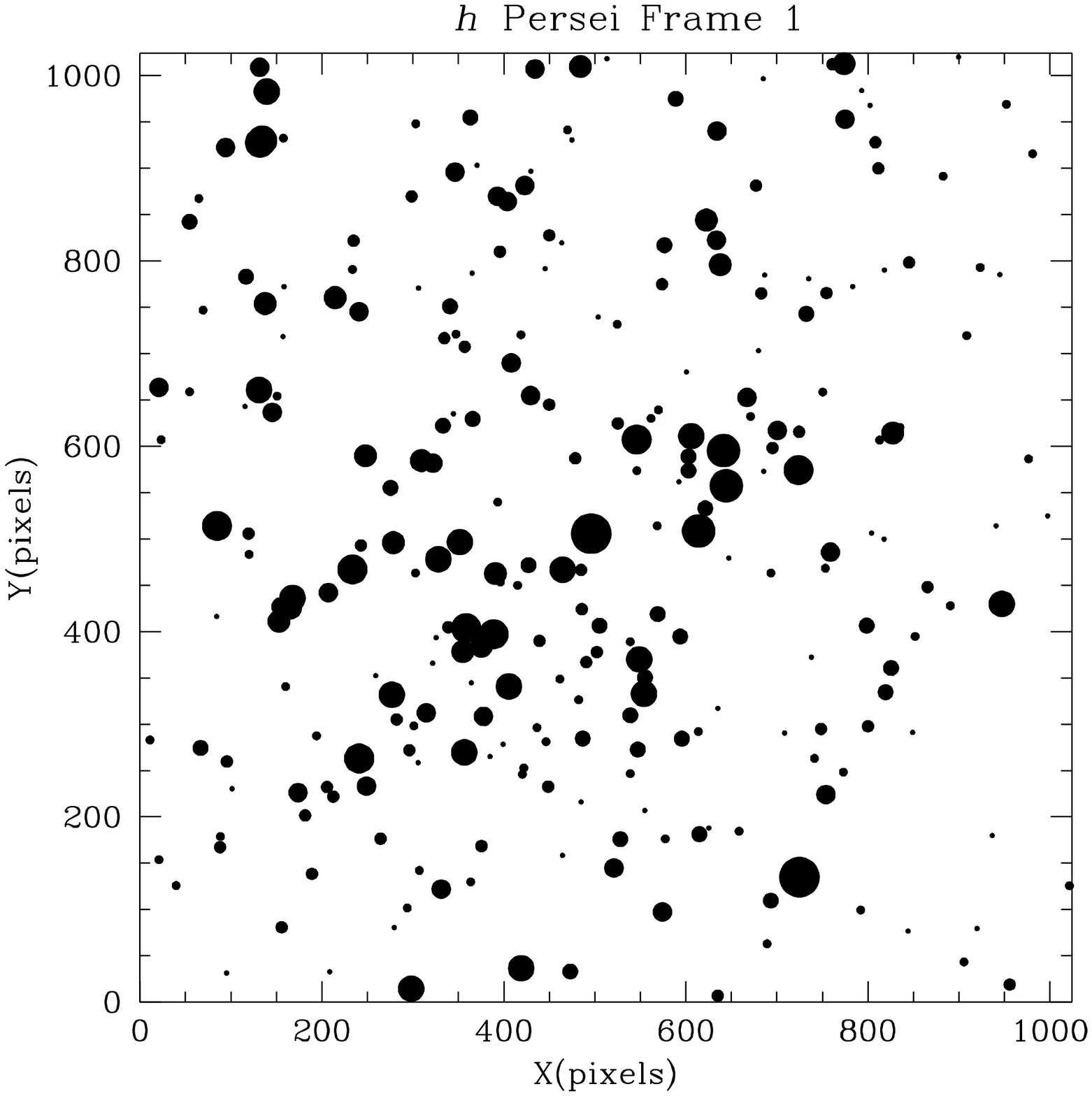,width=8cm,height=8cm} & \psfig{file=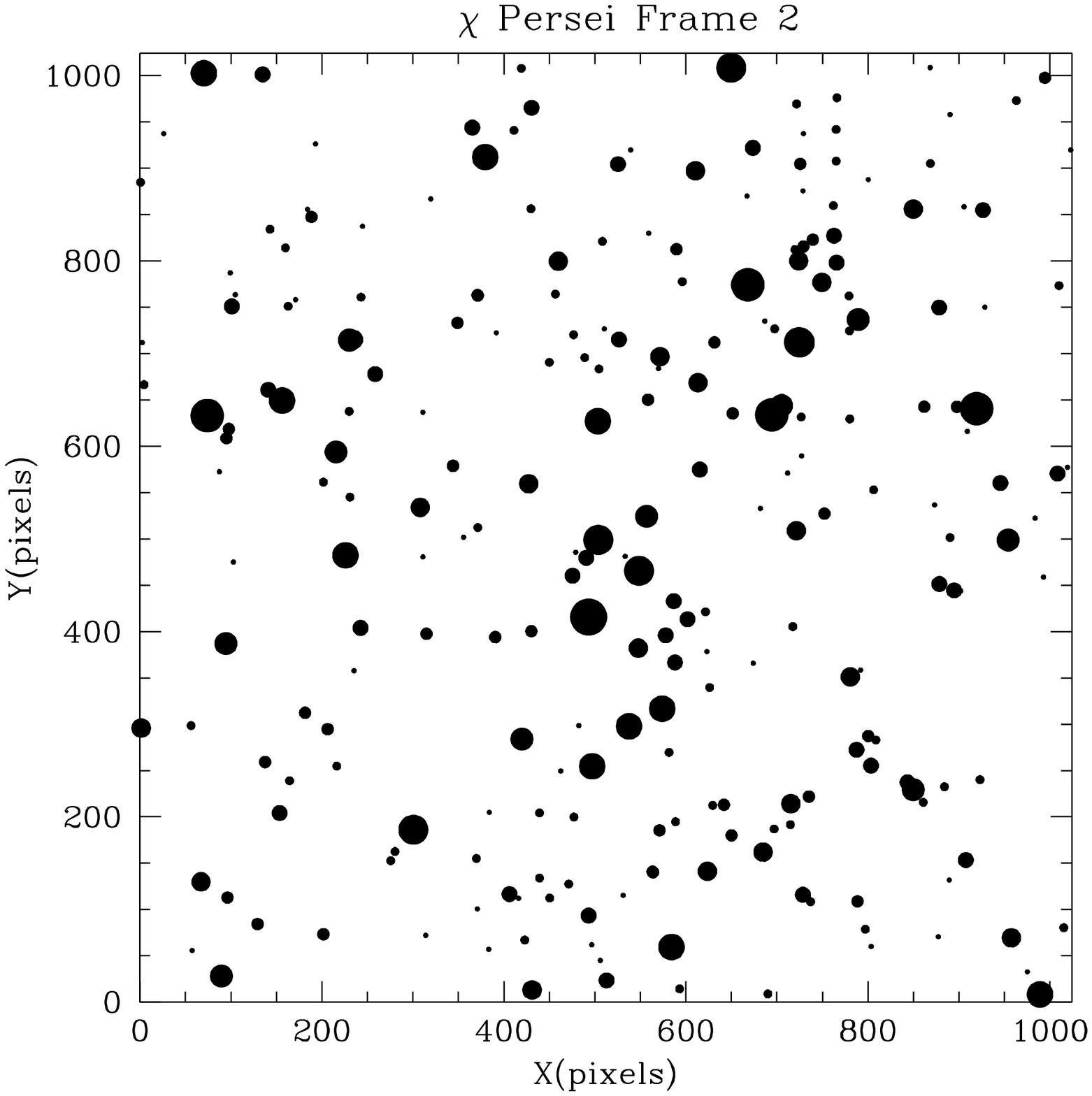,width=8cm,height=8cm}\\
\end{tabular}
\caption{(a)-(b). Schematic maps of the observing region in NGC 884 and NGC 869. The size of the dots represents relative brightness of stars in the field. North is down and East left in all fields.}
\label{fig:coor}
\end{center}
\end{figure*}
The double cluster $h$ \& $\chi$ Persei is one of the richest young open
clusters
accessible from the Northern hemisphere, and therefore it is
well documented in the 
literature. An extensive photographic study was carried out by Oosterhoff
(1937); MK spectral types for cluster members were determined by
Bidelman (1943), Johnson \& Morgan (1955), Schild (1965, 1966, 1967)
and others; optical photometry in various systems was carried out by
Johnson and Morgan (1955), Wildey (1964), Schild (1965), Crawford et al. (1970b), Waelkens et al. (1990), and others; infrared photometry was obtained by Mendoza (1967) and Tapia et al. (1984).
A study of the membership probability of 3086 stars brighter than 
B=15.5 magnitudes within an area of 50 $arcmin$ radius centered on $h$ \&
$\chi$ Persei was carried out by Muminov (1983) on the basis of proper
motion and photometric ($V-($B-V$)$; $V-($U-V$)$; $($U-B$)-($B-V$)$) criteria.
There is no general agreement on the distance moduli and the ages of both clusters. Crawford et al. (1970b) conclude, on the base of $uvby\beta$ photometry, that both clusters have nearly the same age
and distance, the distance modulus being $11.4\pm0.4$ mag.
Balona \& Shobbrook (1984), however, correct this value for evolutionary effects
and adopt a distance modulus of 11.16 for both clusters. On the other hand, Tapia et al. (1984) find confirmation of a previous suggestion by Schild (1967) that $h$ Persei is younger and closer than $\chi$ Persei, with distance moduli
of 11.7 and 12.0, respectively. Doom et al. (1985) show that in the OB
association Per OB1 the low mass stars formed first, the most massive
stars being $(10-20)$ $Myr$ younger than the low mass ones. 
In an attempt to improve these values, we present in this paper deep $uvby\beta$ photometry of
the double open cluster $h$ \& $\chi$ Persei. Given the nearness of these
clusters, our magnitude limit (V=16.5) allows us to reach
 early-F main
sequence stars. This is $3\:{\rm mag}$ deeper than the previous study by
Crawford et al. (1970b) and allows us a very accurate fit to the Zero Age Main 
Sequence (ZAMS) and hence precise determination of the astrophysical 
parameters of the clusters.
\begin{table*}
\caption{Log of the observations taken at the 1.0-m JKT on December 11th and 
December 22nd 1997 for two frames.}
\begin{center}
\begin{tabular}{lccc}
\hline
Cluster&Frame&Central Star& Coordinates (1950)\\
\hline
h Persei&\#1&1057&$\alpha$=2h 15m 32.67s $\delta$=$+56\degr$ $54\farcm$ $20\farcs49$\\
$\chi$ Persei&\#2&2227&$\alpha$=2h 18m 27.70s $\delta$=$+56\degr$ $55\farcm$ $02\farcs08$\\
\hline
Filter&&Exposure Times (s)&\\
&$[V \leq 10]$&&[$V \geq 16$]\\
\hline
$u$&60&300&1200\\
$v$&22&110&450\\
$b$&7&40&150\\
$y$&6&30&120\\
$H\beta_{narrow}$&50&250&1000\\
$H\beta_{wide}$&6&30&120\\
\hline
\hline
\end{tabular}
\end{center}
 \label{tab:coor}
\end{table*}
\section{Observations}

Observations of the central region of $h$ \& $\chi$ Persei were obtained at the 1-m Jacobus Kapteyn Telescope (JKT),
located  at the Observatorio del Roque 
de los Muchachos, La Palma, Spain on the nights of 10--22 December
1997. The telescope was equipped with the 1024 x 1024 TEK 4 chip CCD
and the four Str\"{o}mgren $uvby$ and the narrow and wide H$\beta$
filters. Pixel size was $0\farcs331$ in such a way that the whole
field covered by each frame was $5\farcm6$ x $5\farcm6$. 
Even though both clusters are actually more extended than the area
covered by our frames ($\sim18\arcmin$ across considering the
outermost regions), our images are centered on the central region of
both clusters, where member star density is much higher. 

We took two frames covering the whole of the central region  of
each cluster (central coordinates are displayed in
Table~\ref{tab:coor}). Figures 1(a)-1(b) show plots of the observed
fields in $h$ \& $\chi$ Persei respectively.
The dot sizes are indicative of the relative instrumental $y$ magnitude.
Each cluster was observed using three different exposure times 
in each filter (Table 1),
so that the widest range of magnitudes possible was observed with good 
signal-to-noise ratio. 
Standard stars were observed in the clusters $h$ \& $\chi$ Persei, NGC 6910, 
NGC 2169 and NGC 1039 using the intermediate exposure time.  

Throughout this paper the numbering system used will be that of
Oosterhoff (1937). 
For those stars that were not observed by Oosterhoff (1937), a new
system has been adopted. New stars in $h$ Persei are listed after a ``4'' prefix, 
while new stars in $\chi$ Persei are listed starting with a ``7'' prefix.
Coordinates in each frame for these newly catalogued stars are given in 
Tables~\ref{tab:newcoor1},~\ref{tab:newcoor2} and ~\ref{tab:newcoor3}. 

\begin{table*}
\caption{Pixel coordinates for new (not given by Oosterhoff 1937) stars in the fields of
$h$ and $\chi$ Persei. Number references for stars in the field of $h$ Persei start with
a ``4'' and those for stars in the field of $\chi$ Persei start with a ``7''.}
\begin{center}
\begin{tabular}{ccrrccrr}
\hline
$Nombre$&$Frame$&$X Position$&$Y Position$&$Nombre$&$Frame$&$X Position$&$Y Position$\\
\hline
4000&\#1&693.234&109.661& 7001 &\#2&    0.812  &  884.886\\
4001&\#1&439.143&390.086& 7002 &\#2&    1.464  &  295.878\\
4002&\#1&524.524&731.725&7003 &\#2&    2.551   & 711.971\\
4003&\#1&396.207&453.597&7004 &\#2&    4.584   & 666.505\\
4004&\#1&233.696&790.792&7005 &\#2&   26.156   & 937.495\\
4005&\#1&753.272&468.404&7006 &\#2&   52.413   & 254.809\\
4006&\#1&347.299&721.043&7007 &\#2&   56.227   & 298.508\\
4007&\#1&980.998&915.730&7008 &\#2&   57.494   &  55.824\\
4008&\#1&157.631&932.586&7009 &\#2&   87.365   & 572.511\\
4009&\#1&923.351&793.037&7010 &\#2&   99.222   & 787.185\\
4010&\#1&689.207&63.047&7011 &\#2&  102.705   &  475.030\\
4011&\#1&418.697&720.265&7012 &\#2&  104.895   &  763.610\\
4012&\#1&851.880&394.802&7013 &\#2&  129.354   &   84.270\\
4013&\#1&750.354&658.502&7014 &\#2&  142.978   & 834.276\\
4014&\#1&577.315&176.258&7015 &\#2&   159.980   & 814.161\\
4015&\#1&194.148&287.436&7016 &\#2&  162.952   &  751.130\\
4016&\#1&54.597&658.843&7017 &\#2&  164.448   & 238.962\\
4017&\#1&538.812&388.981&7018 &\#2&  171.039   &  758.260\\
4018&\#1&741.310&263.183&7019 &\#2&  184.205   & 855.733\\
4019&\#1&293.727&101.709&7020 &\#2&  192.982   & 926.343\\
4020&\#1&393.168&539.755&7021 &\#2&  201.651   & 561.368\\
4021&\#1&69.502&746.934&7022 &\#2&  206.334   &  294.670\\
4022&\#1&693.490&463.282&7023 &\#2&  229.956   & 637.612\\
4023&\#1&363.476&129.889&7024 &\#2&  230.787   & 545.157\\
4024&\#1&812.792&606.755&7025 &\#2&  234.797   & 715.244\\
4025&\#1&538.800&246.720&7026 &\#2&  235.338   & 357.776\\
4026&\#1&20.917&153.766&7027 &\#2&  243.004   & 761.009\\
4027&\#1&908.499&719.496&7028 &\#2&  244.684   &  837.470\\
4028&\#1&469.901&941.492&7029 &\#2&   280.240   & 162.642\\
4029&\#1&882.635&891.430&7030 &\#2&  310.902   & 636.701\\
4030&\#1&658.497&184.584&7031 &\#2&  310.965   & 480.615\\
4031&\#1&773.034&248.298&7032 &\#2&  314.016   &  72.096\\
4032&\#1&835.109&620.447&7033 &\#2&  319.664   & 867.014\\
4033&\#1&11.119&283.103&7034 &\#2&  355.690   & 501.842\\
4034&\#1&39.891&125.873&7035  &\#2& 369.775   & 155.062\\
4035&\#1&952.151&969.172&7036  &\#2& 370.781   & 100.616\\
\hline
\end{tabular}
\end{center}
 \label{tab:newcoor1}
\end{table*}

\begin{table*}
\caption{Pixel coordinates for new stars in the fields of
$h$ and $\chi$ Persei (Continued Table 2).}
\begin{center}
\begin{tabular}{ccrrccrr}
\hline
$Nombre$&$Frame$&$X Position$&$Y Position$&$Nombre$&$Frame$&$X Position$&$Y Position$\\
\hline
4036&\#1&1021.319&125.617&7037  &\#2& 371.288   & 512.372\\
4037&\#1&157.298&718.471&7038  &\#2& 371.434   & 762.766\\
4038&\#1&115.477&643.020&7039  &\#2& 383.264   &  56.988\\
4039&\#1&737.855&372.320&7040  &\#2& 383.974   & 204.875\\
4040&\#1&370.525&903.307&7041  &\#2& 391.596   &  722.430\\
4041&\#1&325.514&393.405&7042 &\#2&  416.184   & 112.071\\
4042&\#1&464.297&158.473&7043 &\#2&  419.107   &1008.003\\
4043&\#1&936.568&179.911&7044 &\#2&  422.907   &   67.180\\
4044&\#1&101.268&230.404&7045 &\#2&  429.746   & 856.425\\
4045&\#1&708.475&290.517&7046 &\#2&  439.148   & 204.318\\
4046&\#1&849.082&291.217&7047 &\#2&  439.216   & 133.948\\
4047&\#1&679.772&703.243&7048 &\#2&  450.264   & 112.453\\
4048&\#1&783.255&772.420&7049 &\#2&  462.456   & 249.527\\
4049&\#1&899.591&1020.249&7050 &\#2&  478.848   & 485.563\\
4050&\#1&208.544&32.801&7051 &\#2&  482.343   &  298.590\\
4051&\#1&364.248&344.769&7052 &\#2&  488.737   &   695.700\\
4052&\#1&384.727&265.179&7053 &\#2&  496.502   &   61.990\\
4053&\#1&279.516&80.592&7054 &\#2&   504.380   & 683.503\\
4054&\#1&944.932&785.334&7055 &\#2&  505.889   &  45.079\\
4055&\#1&95.283&31.447&7056 &\#2&  508.301   & 821.225\\
4056&\#1&647.114&479.398&7057 &\#2&  510.341   &  726.720\\
4057&\#1&600.578&680.228&7058 &\#2&  531.088   & 115.350\\
4058&\#1&684.969&996.707&7059 &\#2&  533.233   & 481.181\\
4059&\#1&484.821&216.150&7060 &\#2&  539.252   & 919.885\\
4060&\#1&305.689&258.459&7061 &\#2&  559.066   & 830.051\\
4061&\#1&817.795&499.835&7062 &\#2&  563.855   & 140.253\\
4062&\#1&158.473&772.328&7063 &\#2&   569.820   & 683.924\\
4063&\#1&635.019&317.079&7064 &\#2&  581.503   & 269.484\\
4064&\#1&818.129&790.166&7065 &\#2&  588.605   & 194.763\\
4065&\#1&685.544&572.989&7066 &\#2& 593.093   &  14.331\\
4066&\#1&997.563&524.877&7067 &\#2&  595.982   & 777.736\\
4067&\#1&734.917&780.847&7068 &\#2&  621.566   &421.409\\
4068&\#1&844.050&76.747&7069 &\#2&  623.159   & 378.401\\
4069&\#1&84.295&416.311&7070 &\#2&  626.036   & 339.651\\
4070&\#1&474.699&930.667&7071 &\#2&  629.455   & 212.352\\
\hline
\end{tabular}
\end{center}
 \label{tab:newcoor2}
\end{table*}

\begin{table*}
\caption{Pixel coordinates for new stars in the field of $h$ \&
$\chi$ Persei (Continued Table 2).}
\begin{center}
\begin{tabular}{ccrrccrr}
\hline
$Nombre$&$Frame$&$X Position$&$Y Position$&$Nombre$&$Frame$&
$X Position$&$Y Position$\\
\hline
4071&\#1&803.934&506.322&7072 &\#2&  667.124   & 870.112\\
4072&\#1&321.737&365.810&7073 &\#2&  673.933   & 365.846\\
4073&\#1&919.800&79.571&7074 &\#2&  681.862   & 532.996\\
4074&\#1&364.970&786.852&7075 &\#2&  686.700   & 735.165\\
4075&\#1&429.548&896.836&7076 &\#2&  689.959   &   8.982\\
4076&\#1&445.349&791.666&7077 &\#2&  696.923   & 186.916\\
4077&\#1&306.149&770.777&7078 &\#2&  705.036   & 644.095\\
4078&\#1&592.361&561.733&7079 &\#2&  711.724   & 571.025 \\
4079&\#1&503.583&739.523&7080 &\#2&  714.836   & 191.623\\
4080&\#1&625.128&187.963&7081 &\#2&  717.448   & 405.349\\
4081&\#1&686.471&784.871&7082 &\#2&  718.881   & 218.741\\
4082&\#1&793.110&983.996&7083 &\#2&  719.596   & 812.219\\
4083&\#1&259.138&352.463&7084 &\#2&  721.382   & 508.929\\
4084&\#1&554.851&206.880&7085 &\#2&  721.652   & 969.513\\
4085&\#1&802.316&967.990&7086 &\#2&  725.045   & 712.066\\
4086&\#1&513.151&1018.333&7087 &\#2&  726.728   & 631.566\\
4087&\#1&463.537&819.640&7088 &\#2&  727.113   & 589.622\\
4088&\#1&344.607&635.098&7089 &\#2&  728.574   & 875.621\\
4089&\#1&940.848&514.151&7090&\#2&  729.105   & 937.532\\
4090&\#1&399.125&278.383&7091&\#2&   736.902   & 108.494\\
7092&\#2&   752.186   & 527.338&7093&\#2&    762.070   & 859.814\\
7094&\#2&   765.086   & 941.982&7095&\#2&   765.124   & 907.768\\
7096&\#2&   765.882   & 976.122&7097&\#2&    779.250   & 762.271\\
7098&\#2&  779.759    & 724.668& 7099 &\#2&  779.999   & 629.412\\
7100 &\#2&   791.960   & 358.404& 7101 &\#2&  797.062   &  78.727\\
7102 &\#2&  800.316   & 887.896& 7103 &\#2&  803.498   &  60.048\\
7104 &\#2&  806.325   &553.017& 7105 &\#2&  860.845   & 215.742\\
7106 &\#2&  868.379   &1008.818& 7107 &\#2&  873.146   & 536.758\\
7108 &\#2&  877.309   & 70.644& 7109 &\#2&  884.052   & 232.605\\
7110 &\#2&  889.418   & 131.836& 7111 &\#2&   890.220   & 958.104\\
7112 &\#2&  897.819   & 642.485& 7113 &\#2&  901.955   & 443.717\\
7114 &\#2&  905.646   & 858.522& 7115 &\#2&  909.256   & 616.072\\
7116 &\#2&  923.106   &  240.180& 7117 &\#2&  928.505   & 750.156\\
7118 &\#2&  963.023   & 973.109& 7119 &\#2&  975.222   &  32.738\\
7120 &\#2&  983.717   & 522.477& 7121 &\#2&  992.927   & 458.846\\
7122 &\#2&  994.456   & 997.733& 7123 &\#2& 1009.892   & 773.396\\
7124 &\#2&  1019.220   & 577.266& 7125 &\#2& 1022.753   & 919.784\\
\hline
\end{tabular}
\end{center}
 \label{tab:newcoor3}
\end{table*}
It is worth noting that our sample
is representative of the star population, except in the
sense that it contains almost exclusively main-sequence stars (and
some giants of the earlier spectral types). Most of the brightest
member stars are far away from the central region. This has no bearing
on the determination of the main cluster parameters (such as reddening
and distance), but can influence the study of the system age. For this
reason, we have supplemented our data with observations of a number of
members brighter than $V=11$ taken from Johnson \& Morgan (1955) and 
Crawford et al. (1970b) in Section~\ref{sec:distance}. 

\section{Reduction Procedure}
\label{sec:rdctn}

The reduction of all frames was carried out using the IRAF routines for the bias and 
flat-field corrections. The photometry was obtained by PSF fitting using 
the DAOPHOT package (Stetson 1987) provided by IRAF. The atmospheric 
extinction corrections were performed using the RANBO2 program, which
implements the method described by Manfroid (1993).
It has been shown that the choice of standard stars for the transformation
is a critical issue in $uvby\beta$ photometry. Transformations made
only with unreddened stars  introduce large systematic errors when
applied to reddened stars, even if the colour range of the standards
brackets that of the programme stars (Manfroid \& Sterken 1987;
Crawford 1994). Our data cover a very wide range of spectral types
and hence a wide range of intrinsic 
colours. Moreover, during this campaign several clusters with
different interstellar
reddenings were observed. 

\begin{table*}
\begin{center}
\caption{Standard stars with their catalogued values and spectral types taken from the literature.}
\begin{tabular}{crrrrrc}
\hline
Number&$V$&$b-y$&$m_{1}$&$c_{1}$&$\beta$&$Spectral Type$\\
\hline
\multicolumn{6}{c}{$h$ Persei}\\
\hline
     837&14.080&0.393&0.000&0.918&2.801&          \\
     843& 9.320&0.277&-0.050&0.166&     &B1.5V     \\
     867&10.510&0.393&0.161&0.375&2.613&          \\
     869&      &     &     &     &2.700&          \\
     935&14.020&0.362&-0.004&0.854&2.802&          \\
     950&11.290&0.318&-0.048&0.214&2.642&B2V       \\
     960&      &     &     &     &2.767&          \\
     978&10.590&0.305&-0.039&0.177&2.643&B2V - B1.5V\\
     982&      &     &     &     &2.796&          \\
    1015&10.570&0.225&0.033&0.741&     &B8V       \\
    1078& 9.750&0.316&-0.065&0.167&2.610&B1V - B1Vn\\
    1181&12.650&0.372&-0.034&0.379&2.718&          \\
\hline
\multicolumn{6}{c}{$\chi$ Persei}\\
\hline
    2133&      &     &     &     &2.676&          \\
    2139&11.380&0.255&-0.033&0.196&2.649&B2V       \\
    2147&14.340&0.406&-0.050&1.002&2.863&          \\
    2167&13.360&0.352&-0.056&0.627&2.752&          \\
    2185&10.920&0.283&-0.049&0.406&2.700&B2Vn      \\
    2196&11.570&0.250&-0.006&0.210&2.670&B1.5V     \\
    2200&      &     &     &     &2.721&          \\
    2232&11.110&0.292&-0.105&0.207&2.651&B2V       \\
    2235& 9.360&0.316&-0.088&0.150&2.611&B1V       \\
    2251&11.560&0.302&-0.042&0.349&2.709&B3V       \\
\hline
\multicolumn{6}{c}{$NGC 2169$}\\
\hline
  11&10.600&0.084&0.065&0.541&2.698&B8V       \\
  15&11.080&0.130&0.109&0.944&2.864&B9.5V     \\
  18&11.800&0.115&0.105&0.912&2.872&B9.5V     \\
\hline
\end{tabular}
\end{center}
 \label{tab:estandares}
\end{table*}
\begin{table*}
\begin{center}
\caption{Standard stars with their catalogued values and spectral types taken from the literature.(Continued Table 5).}
\begin{tabular}{crrrrrc}
\hline
Number&$V$&$b-y$&$m_{1}$&$c_{1}$&$\beta$&$Spectral Type$\\
\hline
\multicolumn{6}{c}{$NGC 6910$}\\
\hline
  7&10.360&0.670&-0.160&0.110&2.612&B0.5V     \\
  11&10.900&0.770&0.420&0.430&2.555&          \\
  13&11.720&0.660&-0.140&0.220&2.647&B1V       \\
  14&11.730&0.590&-0.100&0.220&2.652&B1V       \\
  15&12.220&0.590&-0.110&0.330&2.679&          \\
  17&12.660&0.670&-0.120&0.310&2.659&          \\
  18&12.810&0.750&-0.140&0.290&2.680&          \\
  19&12.920&0.640&-0.120&0.380&2.662&          \\
  20&12.980&0.610&-0.130&0.420&2.692&          \\
\hline
\multicolumn{6}{c}{$NGC 1039$}\\
\hline
 92&11.960&0.303&0.138&0.481&2.678&          \\
 96& 9.740&0.086&0.176&0.973&2.890&          \\
 97&11.820&0.144&0.198&0.900&2.855&          \\
 102&10.760&0.151&0.194&0.894&2.848&          \\
 105&11.220&0.176&0.204&0.796&2.817&          \\
 109&10.030&0.066&0.152&1.013&2.916&          \\
 111& 9.950&0.055&0.163&1.021&2.908&          \\
\hline
\end{tabular}
\begin{tabbing}
The data are taken from
Crawford et al. (1970b) \= and Johnson \& Morgan (1955)\\ for $h$ \& $\chi$
Persei, 
Perry et al. (1978) for 
NGC 2169 \=,  Crawford et al. (1977) for NGC 6910 \\ and \= Canterna et al. (1979) for NGC 1039.
Spectral types, when available, are taken from \\ Schild (1965) and Slettebak (1968)
for $h$ \& $\chi$ Persei, Perry et al. (1978) for NGC 2169,\\ Morgan \& Harris (1956) and
Hoag \& Applequist (1965) for NGC 6910 and Canterna \\ et al. (1979) for NGC 1039.\\
\end{tabbing}
\end{center}
 \label{tab:tab2}
\end{table*}

In order to cover the whole range of programme stars, we selected
our standard stars in the same clusters under investigation. A
preliminary list of standard stars
was built by selecting a number of non-variable
non-peculiar candidate stars in 
$h$ \& $\chi$ Persei, NGC 2169, NGC 6910 and NGC 1039,
observed with the same Kitt Peak telescopes and instrumentation used to
define the $uvby$ Crawford \& Barnes (1970a) and H$\beta$ Crawford \&
Mander (1966) standard systems, so that there is no doubt that the 
photometric values are in the standard systems. Since the original 
observations of $h$
\& $\chi$ Persei by Crawford et al. (1970b) do not include
$V$ values, we used values given by Johnson \& Morgan (1955), which
were also taken with the same instrumentation,
for the $V$ transformation.

The list of adopted standard stars and their photometric data to be
used in the transformations are given in Tables 5 and 6.

\begin{table*}[!]
\begin{center}
\caption{Catalogue of 41 standard stars observed and transformed to the Crawford-Barnes $uvby$ and the Crawford-Mander $H\beta$ standard systems. The internal rms errors of the mean measure in the transformation of each star are given in columns 8 to 11 in units of 0.001 mag. Columns 12 to 16 give the difference D=standard value minus transformed value, in units of 0.001 mag. N is the number of measures of each standard star in the transformation in $V$, $(b-y)$, $m_{1}$, $c_{1}$. The number of measures in $\beta$ is one for each standard star in the transformation.}
\begin{tabular}{lrrrrc@{}crrrrrrrrr}
\hline
Star&V&$(b-y)$&$m_{1}$&$c_{1}$&$\beta$&$N_{uvby}$&$\sigma_{V}$&$\sigma_{b-y}$&$\sigma_{m_{1}}$&$\sigma_{c_{1}}$&$D_{V}$&$D_{(b-y)}$&$D_{m_{1}}$&$D_{c_{1}}$&$D_{\beta}$\\
\hline
\multicolumn{15}{c}{$h$ Persei}\\
\hline
869&-&-&-&-&2.724&-&-&-&-&-&-&-&-&-&-024\\
837& 14.095&0.411&-0.031&0.907&2.789& 3&006&009&024& 024&-016&-019& 032&011&012\\
843&  9.330&0.319&-0.151&0.240&-& 1&  -   &  -   &  -   &  -    &-010&-042& 101&-074&-\\
867& 10.572&0.376& 0.177&0.379&2.666& 5&006&008&013& 025&-063& 016&-015&-005&-053\\
935& 14.051&0.411&-0.076&0.856&2.781& 6&018&007&018& 038&-031&-049& 073&-003&021\\
950& 11.297&0.337&-0.079&0.221&2.630& 1&  -   &  -   &  -   &  -  &-007&-019&031&-007&012\\
960&-&-&-&-&2.727&-&-&-&-&-&-&-&-&-&040\\
978& 10.646&0.328&-0.070&0.194&2.634& 5&016&013&016&023&-056&-023& 031&-017&009\\
982&-&-&-&-&2.779&-&-&-&-&-&-&-&-&-&017\\
1015& 10.573&0.223& 0.029&0.677&-& 2&037&023&006& 030&-003& 003& 005&064&-\\
1078&  9.775&0.317&-0.043&0.138&2.611& 3&008&021&042& 036&-025&-002&-021&030&-001\\
1181& 12.655&0.352&-0.008&0.338&2.703& 6&022&022&044& 030&-006& 020&-025&041&015\\
\hline
\multicolumn{15}{c}{$\chi$ Persei}\\
\hline
2133&-&-&-&-&2.658&-&-&-&-&-&-&-&-&-&018\\
2139& 11.351&0.298&-0.097&0.235&2.641& 2&011&024&030& 007&029&-043& 064&-039&008\\
2147& 14.359&0.392&-0.083&1.022&2.783& 5&023&034&051& 051&-019& 013& 033&-020&080\\
2167& 13.364&0.347&-0.080&0.616&2.733& 6&012&008&012& 024&-004& 005& 025&011&019\\
2185& 10.926&0.275&-0.018&0.412&2.688& 3&008&013&038& 044&-006& 008&-031&-006&012\\
2196& 11.549&0.304&-0.066&0.246&2.632& 6&012&013&023&023&021&-052& 056&-034&038\\
2200&-&-&-&-&2.707&-&-&-&-&-&-&-&-&-&014\\
2232& 11.052&0.238&-0.029&0.177&2.639& 4&013&013&041& 035&058& 054&-077&030&012\\
2235&  9.365&0.311&-0.071&0.131&2.575& 3&012&015&046& 039&-005& 005&-016&018&036\\
2251& 11.563&0.297&-0.028&0.367&2.689& 6&009&011&027& 030&-004& 004&-013&-018&020\\
\hline
\multicolumn{15}{c}{$NGC 2169$}\\
\hline
11& 10.538&0.076& 0.061&0.545&2.719& 1&-     &-     &   -  &  -   &062& 008& 004&-004&-021\\
15& 11.023&0.122& 0.092&0.939&2.856& 1&-     &   -  &   -  & -    &057& 008& 017&005&008\\
18& 11.733&0.053& 0.260&0.813&2.883& 1&-     &-     & -    &   -  &067& 062&-155&099&-011\\
\hline
\end{tabular}
\end{center}
 \label{tab:standtrans}
\end{table*}
\begin{table*}[!]
\begin{center}
\caption{Catalogue of 41 standard stars observed and transformed to the Crawford-Barnes $uvby$ and the Crawford-Mander $H\beta$ standard systems. The internal rms errors of the mean measure in the transformation of each star are given in columns 8 to 11 in units of 0.001 mag. Columns 12 to 16 give the difference D=standard value minus transformed value, in units of 0.001 mag. N is the number of measures of each standard star in the transformation in $V$, $(b-y)$, $m_{1}$, $c_{1}$. The number of measures in $\beta$ is one for each standard star in the transformation. (Continued Table 7).}
\begin{tabular}{lrrrrc@{}crrrrrrrrr}
\hline
Star&V&$(b-y)$&$m_{1}$&$c_{1}$&$\beta$&$N_{uvby}$&$\sigma_{V}$&$\sigma_{b-y}$&$\sigma_{m_{1}}$&$\sigma_{c_{1}}$&$D_{V}$&$D_{(b-y)}$&$D_{m_{1}}$&$D_{c_{1}}$&$D_{\beta}$\\
\hline
\multicolumn{15}{c}{$NGC 6910$}\\
\hline
7& 10.320&0.667&-0.150&0.092&2.652& 2&011&004&012& 016&041& 003&-011&018&-040\\
11&-&-&-&-&2.639&-&-&-&-&-&-&-&-&-&-084\\
13& 11.681&0.619&-0.082&0.211&2.661& 2&004&023&038& 004&039& 041&-058&009&-014\\
14& 11.730&0.569&-0.053&0.201&2.678& 2&004&028&040& 018&001& 021&-048&019&-026\\
15& 12.193&0.609&-0.134&0.359&2.698& 2&007&006&014& 002&027&-019& 024&-029&-019\\
17& 12.635&0.700&-0.149&0.294&2.688& 2&008&010&025& 040&025&-030& 029&017&-029\\
18& 12.816&0.755&-0.132&0.296&2.705& 2&014&004&006& 048&-006&-005&-009&-006&-025\\
19& 12.897&0.610&-0.065&0.402&2.709& 2&025&035&037& 047&023& 031&-056&-022&-047\\
20& 12.920&0.619&-0.118&0.446&2.730& 2&016&001&019& 006&060&-009&-013&-026&-038\\
\hline
\multicolumn{15}{c}{$NGC 1039$}\\
\hline
92& 11.928&0.282& 0.150&0.520&2.709& 1&  -   &  -   &  -   &  -   &032& 021&-012&-039&-031\\
96&  9.703&0.063& 0.219&0.976&2.887& 1&-     &  -   &  -   &   -  &037& 023&-043&-003&003\\
97& 11.782&0.129& 0.215&0.934&2.838& 1&-     &  -   &  -   &  -   &038& 015&-017&-034&017\\
102& 10.724&0.150& 0.193&0.893&2.851& 1&-     & -    &  -   &    - &036& 001& 001&001&-003\\
105& 11.166&0.167& 0.232&0.971&2.832& 1&-     & -    & -    &   -  &054& 009& 028&-175&-015\\
109& 10.034&0.040& 0.178&1.018&2.960& 1&-     &-     & -    & -    &-004& 026&-026&-005&-044\\
111&  9.900&0.025& 0.249&0.964&2.927& 1& -    & -    &  -   &  -   &050& 030&-086&057&-019\\
\hline
\end{tabular}
\end{center}
 \label{tab:tab3}
\end{table*}

The following $uvby$ transformation equations from the instrumental
to the standard system together with the standard errors on the coefficients are obtained using the equations by Crawford \& Barnes (1970a), where the coefficients have been computed following the procedure described in detail by Gr\o nbech et al. (1976):

\begin{equation}
\begin{alignat}{2}
V = \,&11.269\, &+\, &0.091(b-y)\,+\,y_{{\rm i}}\\
&\pm0.003& &\pm0.007 \nonumber
\end{alignat}
\end{equation}
\begin{equation}
\begin{alignat}{2}
(b-y) = \,&0.636\, & +\, &1.070(b-y)_{{\rm i}}\\
&\pm0.008 & &\pm0.008 \nonumber
\end{alignat}
\end{equation}
\begin{equation}
\begin{alignat}{3}
m_{1} =\, &-0.523\, &+\, &1.009m_{1{\rm i}} &\,-\, &0.206(b-y)\\
&\pm0.008 & &\pm0.015 & &\pm0.009   \nonumber
\end{alignat}
\end{equation}
\begin{equation}
\begin{alignat}{3}
c_{1} =\, &0.543\, &+\, &1.019c_{1{\rm i}} &\,+\, &0.257(b-y)\\
&\pm0.004 & & \pm0.004 & &\pm0.007 \nonumber
\end{alignat}
\end{equation}  
where the index ``i'' stands for instrumental magnitudes.

The transformed values for
the 41 standard stars are given in Tables 7 and 8, together
with their precision and deviation with respect to the published standard
values.
Table 9 shows
the mean catalogue minus transformed values for the standard
stars and their standard deviations, which constitute a measure of the accuracy of 
the transformation. From the mean differences between catalogue and transformed 
values, it is clear that there is not a significant offset between our photometry
and the standard system. Since the individual differences for a few stars seem
to be rather large, an attempt was made to improve the transformation by removing
these stars from the standard list. We find however that the transformation
coefficients and their precision do not improve significantly. 

\begin{table*}[t]
\begin{center}
\caption{Mean catalogue minus transformed values for the standard
stars and their standard deviations in $V$, $(b-y)$, $m_{1}$, $c_{1}$ and $\beta$.}
\begin{tabular}{rrrrr}
\hline
$D_{m{V}}$&$D_{m{(b-y)}}$&$D_{m{m_{1}}}$&$D_{m{c_{1}}}$&$D_{m{\beta}}$\\
\hline
0.014&0.003&-0.005&-0.004&-0.003\\
0.033&0.027&0.049&0.044&0.031\\
\hline
\hline
\end{tabular}
\end{center}
 \label{tab:tab4}
\end{table*}

The H$\beta$ instrumental system and transformation equations were computed following
the procedure described in detail by Crawford \& Mander (1966). The transformation
coefficients are $a=3.514$ and $b=1.059$. Transformed values and their differences
with respect to the mean catalogue values are given in Tables 7 and 8. 
The mean difference is $-0.003$ with a standard deviation $0.031$, which, as in the
case of the $uvby$ transformation, indicates that there is no significant offset
with respect to the standard system. 

\section{Results}

\subsection{Membership and Reddening}

We have obtained $uvby\beta$ CCD photometry for more than 350 stars 
in the fields of $h$ \& $\chi$ Persei. The magnitude limit $V\approx 16.5$ 
allows us to identify 214 stars that were not catalogued by Oosterhoff (1937).
Even though all of them are listed in Tables 2, 3 and 4 some of 
them are so faint that the number of counts was not enough to 
reach a
good signal-to-noise ratio in all the filters. Therefore these stars
are discarded from our sample.

\begin{figure}
\begin{picture}(240,300)
\put(0,0){\includegraphics{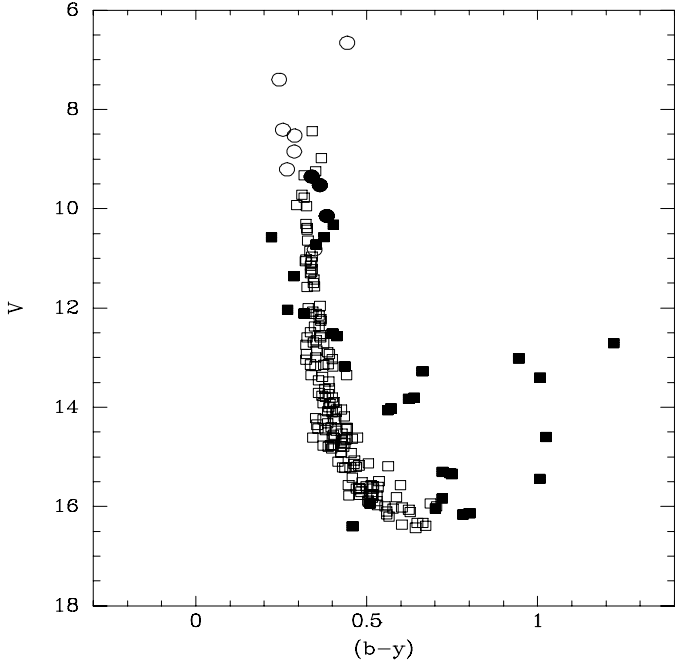}}
\end{picture}
\caption{$V$ -- $(b-y)$ diagram for all stars in the field of 
$h$ Persei. Open squares represent stars considered as members while filled squares are non-members. Filled circles are stars catalogued as Be stars. Open circles are supergiant and giant
stars not observed by us and taken from the study of Crawford et al. (1970b).}
\label{fig:Vb-yh}
\end{figure}
\begin{figure}
\begin{picture}(240,300)
\put(0,0){\includegraphics{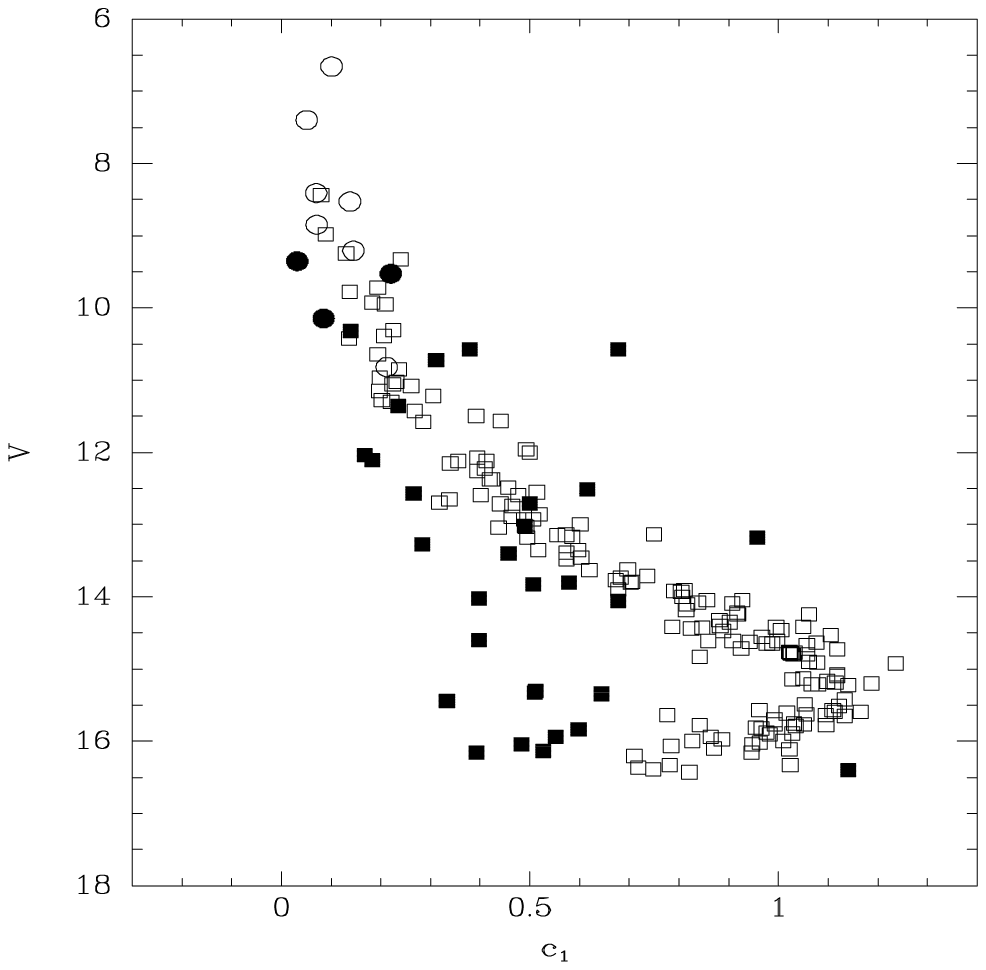}}
\end{picture}
\caption{$V$ -- $c_{1}$ diagram for all stars in the field of 
$h$ Persei. Open squares represent stars considered as members while filled squares are non-members. Filled circles are stars catalogued as Be stars. Open circles are supergiant and giant
stars not observed by us and taken from the study of Crawford et al. (1970b).}
\label{fig:Vc1h}
\end{figure}

\begin{figure}
\begin{picture}(240,300)
\put(0,0){\includegraphics{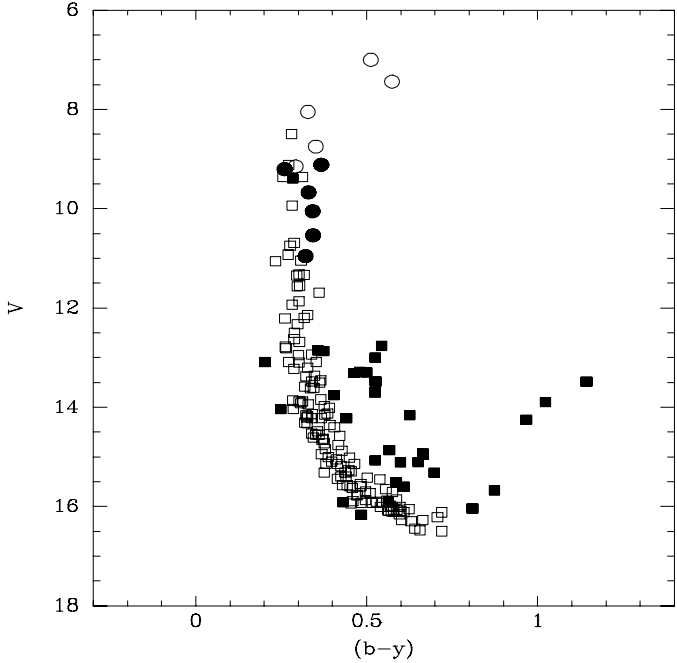}}
\end{picture}
\caption{$V - (b-y)$ diagram for all stars in the field of 
$\chi$ Persei. The open squares are considered as members and the filled squares as non-members. The filled circles are stars catalogued as Be stars. Open circles are supergiant and giant
stars not observed by us and taken from the study of Crawford et al. (1970b).}
\label{fig:Vb-ychi}
\end{figure}

\begin{figure}
\begin{picture}(240,300)
\put(0,0){\includegraphics{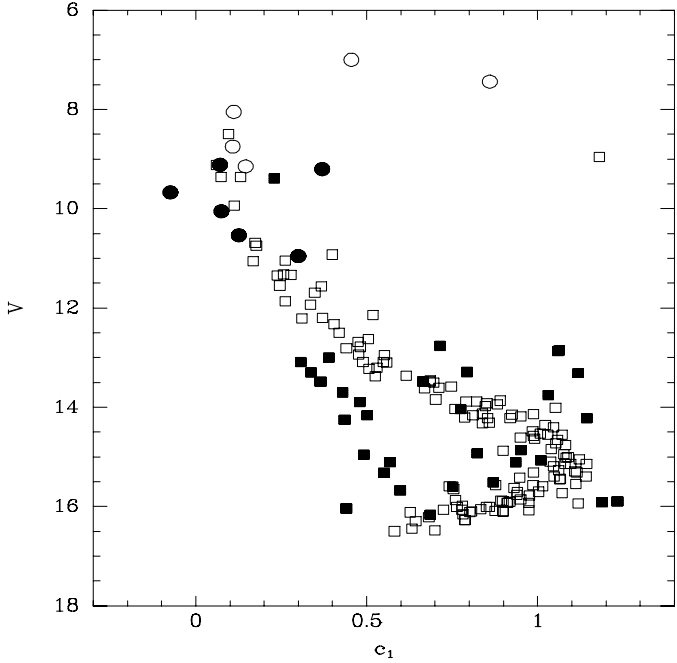}}
\end{picture}
\caption{$V$ -- $c_{1}$ diagram for all stars in the field of 
$\chi$ Persei. The open squares are considered as members and the filled squares as non-members. The filled circles are stars catalogued as Be stars. Open circles are supergiant and giant
stars not observed by us and taken from the study of Crawford et al. (1970b).}
\label{fig:Vc1chi}
\end{figure}

To assess the membership of a star, we look at its position in the $V-(b-y)$
and $V-c_{1}$ diagrams. 
We find that in both diagrams the vast majority of the stars fall along 
a very well
defined main sequence.
Inspection of these photometric diagrams reveals that a number
of stars do not fit well the
main sequence loci in both diagrams. Those objects are considered
as non-members unless 
they are catalogued as Be stars (see Figures~\ref{fig:Vb-yh},
\ref{fig:Vc1h}, \ref{fig:Vb-ychi} and \ref{fig:Vc1chi}).
Indeed Be stars have colours differing 
from those of non-emission B stars due to additional reddening 
caused by the circumstellar envelope and tend to
have redder $(b-y)$ and lower $c_{1}$ values than normal B stars 
(Fabregat et al. 1996). In addition, the M4.5Iab supergiant star
RS Per (2417) 
is considered to be a cluster member. 

\begin{table*}
\begin{center}
\caption{Photometric data for members of $h$ Persei.}
\begin{tabular}{crcrccccccccc}
\hline
Number&$V$&$b-y$&$m_{1}$&$c_{1}$&$\beta$&$\sigma_{V}$&$\sigma_{b-y}$&$\sigma_{m_{1}}$&$\sigma_{c_{1}}$&$\sigma_{\beta}$&$N_{uvby}$&$N_{\beta}$\\
\hline
 820&13.173& 0.350&-0.077& 0.585& 2.720&0.010& 0.012& 0.030&0.045&  -  & 2&  1\\
 821&15.492& 0.536&-0.056& 1.053& 2.806& -   &  -   &  -   &  -  &  -  & 1&  1\\
 832&14.217& 0.351&-0.006& 0.917& 2.839&0.002& 0.002& 0.004&0.045&     & 2&  1\\
 836&15.775& 0.448& 0.121& 0.842& 2.852& -   &  -   &  -   &  -  &  -  & 1&  1\\
 837&14.095& 0.411&-0.031& 0.907& 2.789&0.006& 0.009& 0.024&0.024&     & 3&  1\\
 842&13.142& 0.335&-0.051& 0.573& 2.747&0.005& 0.015& 0.026&0.035&0.006& 6&  3\\
 843& 9.330& 0.318&-0.151& 0.240&  -   & -   &  -   &  -   &  -  &  -  & 1&  0\\
 844&15.592& 0.499&-0.035& 1.113& 2.821&0.011& 0.047& 0.086&0.019&  -  & 3&  1\\
 845&14.764& 0.394&-0.038& 1.022& 2.880&0.012& 0.023& 0.058&0.072&0.032& 4&  2\\
 848&15.095& 0.416&-0.005& 1.118& 2.882&0.008& 0.012& 0.031&0.053&0.004& 3&  2\\
 854&14.240& 0.405&-0.006& 1.061& 2.910&0.014& 0.026& 0.044&0.054&0.006& 6&  2\\
 856&15.591& 0.518&-0.101& 1.165& 2.860&0.004& 0.016& 0.056&0.094&0.030& 3&  2\\
 857&14.909& 0.425&-0.034& 1.078& 2.903&0.008& 0.026& 0.032&0.047&0.036& 4&  2\\
 864& 9.927& 0.295&-0.049& 0.183& 2.648&0.012& 0.001& 0.006&0.017&0.012& 3&  2\\
 869&12.002&0.329&-0.065&0.500&2.714&0.012&0.005&0.009&0.016&0.012&6&3\\
 875&15.611& 0.533&-0.112& 1.017& 2.896&0.011& 0.003& 0.021&0.084&  -  & 2&  1\\
 876&12.715& 0.374&-0.132& 0.440& 2.703&0.014& 0.013& 0.027&0.028&0.008& 6&  3\\
 879&11.578& 0.326&-0.116& 0.285& 2.648&0.010& 0.009& 0.029&0.043&0.011& 6&  3\\
 880&13.029& 0.399&-0.147& 0.492& 2.671&0.011& 0.020& 0.044&0.059&0.008& 6&  3\\
 885&15.762& 0.478& 0.018& 1.051& 2.817&0.011& 0.011& 0.020&0.052&  -  & 3&  1\\
 892&11.216& 0.340&-0.150& 0.305& 2.649&0.008& 0.009& 0.030&0.045&0.022& 5&  2\\
 893&11.964& 0.363&-0.116& 0.492& 2.697&0.013& 0.007& 0.029&0.059&0.014& 6&  3\\
 896&12.489& 0.335&-0.061& 0.456& 2.678&0.014& 0.013& 0.023&0.031&0.020& 6&  3\\
 898&14.552& 0.403&-0.041& 0.967& 2.834&0.011& 0.014& 0.030&0.060&0.007& 4&  2\\
 901&14.902& 0.425&-0.041& 1.061& 2.854&0.002& 0.019& 0.036&0.029&0.013& 4&  2\\
 907&12.374& 0.348&-0.119& 0.419& 2.687&0.012& 0.015& 0.032&0.039&0.018& 6&  3\\
 909&15.188& 0.563&-0.023& 1.114& 2.856&0.017& 0.022& 0.057&0.101&0.036& 3&  2\\
 914&15.202& 0.470&-0.039& 1.187& 2.875&0.011& 0.020& 0.061&0.082&0.026& 3&  2\\
 917&14.775& 0.373& 0.036& 1.023& 2.904&0.016& 0.017& 0.032&0.027&0.058& 4&  2\\
 923&13.044& 0.323&-0.045& 0.437& 2.711&0.011& 0.015& 0.029&0.023&0.018& 6&  3\\
 924&15.173& 0.477&-0.104& 1.098& 2.846&0.017& 0.050& 0.094&0.090&0.010& 3&  2\\
 926&11.570&0.347&-0.085&0.441&2.689&0.011&0.011&0.019&0.019&0.004&6&3\\
 929&10.306& 0.322&-0.089& 0.225& 2.625&0.015& 0.009& 0.027&0.033&0.004& 4&  2\\
 930&12.377& 0.361&-0.051& 0.424& 2.708&0.012& 0.010& 0.019&0.024&0.007& 6&  3\\
 934&14.645& 0.440&-0.036& 0.977& 2.830&0.029& 0.040& 0.058&0.063&0.027& 5&  2\\
 935&14.051& 0.411&-0.076& 0.856& 2.771&0.018& 0.007& 0.018&0.038&0.013& 6&  3\\ 
\hline
\end{tabular}
\end{center}
 \label{tab:miembrosh1}
\end{table*}

\begin{table*}
\begin{center}
\caption{Photometric data for members of $h$ Persei (Continued Table 10).}
\begin{tabular}{crcrccccccccc}
\hline
Number&$V$&$b-y$&$m_{1}$&$c_{1}$&$\beta$&$\sigma_{V}$&$\sigma_{b-y}$&$\sigma_{m_{1}}$&$\sigma_{c_{1}}$&$\sigma_{\beta}$&$N_{uvby}$&$N_{\beta}$\\
\hline
 936&10.395& 0.325&-0.122& 0.206& 2.625&0.019& 0.011& 0.030&0.042&0.003& 4&  2\\
 939&12.263& 0.366&-0.130& 0.394& 2.688&0.013& 0.017& 0.033&0.034&0.020& 6&  3\\
 941&15.217& 0.429& 0.000& 1.067& 2.868&0.013& 0.020& 0.026&0.053&0.018& 3&  2\\
 945&14.796& 0.387& 0.019& 1.030& 2.893&0.012& 0.008& 0.017&0.042&0.004& 4&  2\\
 946&14.418& 0.439&-0.013& 1.050& 2.840&0.022& 0.023& 0.042&0.055&0.028& 6&  2\\
 947&15.828& 0.517& 0.068& 0.966& 2.853&0.002& 0.017& 0.075&0.058&  -  & 2&  1\\
 948&15.633& 0.477& 0.011& 1.056& 2.861&0.025& 0.029& 0.049&0.029&0.052& 3&  2\\
 949&15.704& 0.483& 0.012& 0.992& 2.901&0.004& 0.006& 0.010&0.012&  -  & 2&  1\\
 950&11.297& 0.336&-0.079& 0.221& 2.630& -   &  -   &  -   &   - &  -  & 1&  1\\
 952&12.072& 0.342&-0.072& 0.394& 2.674&0.012& 0.009& 0.014&0.012&0.006& 6&  3\\
 955&15.781& 0.530&-0.017& 1.096& 2.814&0.020& 0.044& 0.088&0.074&  -  & 3&  1\\
 956&12.554& 0.363&-0.085& 0.514& 2.704&0.017& 0.017& 0.041&0.049&0.005& 6&  3\\
 959&12.858& 0.354&-0.068& 0.519& 2.702&0.012& 0.013& 0.023&0.028&0.021& 6&  3\\
 960&13.715& 0.360&-0.021& 0.736& 2.743&0.014& 0.024& 0.046&0.041&0.014& 6&  3\\
 963&11.021& 0.323&-0.116& 0.232& 2.638&0.009& 0.007& 0.012&0.020&0.001& 5&  2\\
 965&12.595& 0.365&-0.057& 0.476& 2.709&0.009& 0.009& 0.019&0.032&0.029& 6&  3\\
 966&14.246& 0.372&-0.001& 0.918& 2.851&0.016& 0.023& 0.057&0.028&0.023& 6&  2\\
 970&14.403& 0.396&-0.086& 0.883& 2.788&0.041& 0.055& 0.087&0.053&0.001& 5&  2\\
 971&14.460& 0.379&-0.012& 1.005& 2.839&0.036& 0.050& 0.076&0.036&0.013& 6&  2\\
 976&11.498&0.344&-0.081&0.392&2.689&0.010&0.012&0.022&0.032&0.013&6&3\\
 978&10.646& 0.328&-0.070& 0.194& 2.632&0.016& 0.013& 0.016&0.023&0.003& 5&  2\\
 979&14.246& 0.409& 0.045& 0.917& 2.886&0.019& 0.016& 0.026&0.034&0.016& 6&  2\\
 980& 9.716& 0.310&-0.082& 0.194& 2.621&0.029& 0.018& 0.016&0.020&0.009& 3&  2\\
 982&13.919& 0.372&-0.037& 0.790& 2.780&0.012& 0.015& 0.038&0.041&0.027& 6&  3\\
 985&12.118& 0.337&-0.075& 0.413& 2.691&0.013& 0.017& 0.026&0.035&0.006& 6&  3\\
 986&12.589& 0.325&-0.042& 0.401& 2.684&0.018& 0.013& 0.021&0.028&0.004& 6&  3\\
 987&14.635& 0.458&-0.012& 1.076& 2.899&0.033& 0.036& 0.034&0.028&0.022& 5&  2\\
 988&12.749& 0.322&-0.005& 0.465& 2.711&0.014& 0.011& 0.023&0.024&0.009& 6&  3\\
 990&14.429& 0.357& 0.023& 0.847& 2.833&0.005& 0.025& 0.043&0.050&0.033& 5&  2\\
 991&11.427& 0.346&-0.086& 0.268& 2.641&0.016& 0.012& 0.020&0.013&0.005& 6&  3\\
 992& 9.952& 0.324&-0.091& 0.209& 2.641&0.008& 0.012& 0.017&0.002&  -  & 3&  1\\
 997&11.086& 0.338&-0.097& 0.261& 2.651&0.011& 0.006& 0.006&0.012&0.006& 5&  2\\
 999&13.354& 0.338& 0.001& 0.597& 2.736&0.017& 0.019& 0.034&0.037&0.012& 6&  3\\
1000&13.140&0.387&-0.024&0.750&2.790&0.018&0.015&0.025&0.019&0.010&6&3\\
1004&10.849& 0.334&-0.089& 0.237& 2.638&0.012& 0.005& 0.013&0.025&0.004& 5&  2\\
1007&15.575& 0.447& 0.067& 1.109& 2.865&0.010& 0.039& 0.070&0.150&  -  & 3&  1\\
\hline
\end{tabular}
\end{center}
 \label{tab:miembrosh2}
\end{table*}

\begin{table*}
\begin{center}
\caption{Photometric data for members of $h$ Persei (Continued Table 10).}
\begin{tabular}{crcrccccccccc}
\hline
Number&$V$&$b-y$&$m_{1}$&$c_{1}$&$\beta$&$\sigma_{V}$&$\sigma_{b-y}$&$\sigma_{m_{1}}$&$\sigma_{c_{1}}$&$\sigma_{\beta}$&$N_{uvby}$&$N_{\beta}$\\
\hline
1014&12.926& 0.324& 0.007& 0.488& 2.737&0.018& 0.022& 0.042&0.036&0.026& 6&  3\\
1017&15.791& 0.514& 0.026& 1.035& 2.791&0.015& 0.027& 0.032&0.052&  -  & 3&  1\\
1018&15.653& 0.479& 0.017& 1.134& 2.884&0.022& 0.022& 0.019&0.034&  -  & 3&  1\\
1020&13.771& 0.370&-0.047& 0.673& 2.700&0.019& 0.012& 0.024&0.028&0.031& 6&  3\\
1021&12.997& 0.351&-0.020& 0.601& 2.762&0.016& 0.013& 0.020&0.020&0.019& 6&  3\\
1025&15.751& 0.515& 0.006& 1.031& 2.890&  -  &   -  &  -   &  -  &  -  & 1&  1\\
1028&14.778& 0.401& 0.028& 1.032& 2.863&0.014& 0.031& 0.072&0.022&0.003& 3&  2\\
1030&16.048& 0.577& 0.011& 0.947& 2.835&  -  &   -  &  -   &  -  &  -  & 1&  1\\
1031&15.216& 0.452& 0.025& 1.080& 2.886&0.024& 0.044& 0.064&0.012&0.037& 3&  2\\
1034&15.146& 0.463& 0.006& 1.028& 2.881&0.009& 0.022& 0.033&0.012&0.013& 3&  2\\
1038&15.570& 0.513&-0.017& 1.109& 2.867&0.029& 0.037& 0.052&0.020&  -  & 3&  1\\
1041&11.057& 0.321&-0.054& 0.224& 2.645&0.014& 0.014& 0.019&0.009&0.016& 5&  2\\
1049&13.897& 0.404&-0.084& 0.678& 2.746&0.027& 0.018& 0.034&0.047&  -  & 4&  1\\
1050&16.002& 0.553&-0.010& 1.009& 2.843&  -  &   -  &   -  &  -  &  -  & 1&  1\\
1052&15.226& 0.436&-0.007& 1.140& 2.906&0.024& 0.026& 0.043&0.054&0.032& 3&  2\\
1053&14.613& 0.342& 0.076& 0.859& 2.916&0.038& 0.038& 0.081&0.051&0.002& 5&  2\\
1056&14.671& 0.428&-0.038& 1.057& 2.860&0.015& 0.022& 0.025&0.029&0.027& 4&  2\\
1058&13.621& 0.390&-0.054& 0.697& 2.726&0.023& 0.024& 0.055&0.070&0.025& 6&  3\\
1059&14.728& 0.435&-0.041& 1.119& 2.880&0.024& 0.016& 0.018&0.058&0.081& 4&  2\\
1064&14.613& 0.400&-0.007& 0.908& 2.862&0.013& 0.019& 0.036&0.045&0.034& 4&  3\\
1066&13.145& 0.374&-0.050& 0.555& 2.728&0.013& 0.021& 0.038&0.028&0.013& 6&  3\\
1072&13.353&0.441&-0.057&0.517&2.693&0.015&0.016&0.025&0.019&0.015&6&3\\
1077&13.792& 0.379&-0.010& 0.706& 2.764&0.020& 0.015& 0.032&0.038&0.031& 6&  3\\
1078& 9.775& 0.317&-0.043& 0.138& 2.610&0.008& 0.021& 0.042&0.036&0.001& 3&  2\\
1079&15.639& 0.470& 0.067& 0.777& 2.835&0.040& 0.000& 0.026&0.080&  -  & 2&  1\\
1080&11.151& 0.336&-0.047& 0.197& 2.644&0.015& 0.014& 0.025&0.018&0.004& 5&  2\\
1081&14.320& 0.386&-0.026& 0.881& 2.771&0.047& 0.031& 0.026&0.023&0.019& 5&  2\\
1083&13.388& 0.370&-0.029& 0.574& 2.734&0.021& 0.015& 0.026&0.029&0.017& 6&  3\\
1085&10.426& 0.326&-0.051& 0.136& 2.625&0.014& 0.014& 0.018&0.010&0.003& 5&  2\\
1091&15.878& 0.504& 0.050& 0.976& 2.840&0.006& 0.018& 0.057&0.100&  -  & 2&  1\\
1093&13.486& 0.389&-0.058& 0.574& 2.719&0.025& 0.020& 0.029&0.017&0.022& 6&  3\\
1095&13.635& 0.378& 0.005& 0.620& 2.724&0.014& 0.018& 0.050&0.041&0.003& 6&  3\\
1096&15.078& 0.464& 0.019& 1.119& 2.942&0.005& 0.020& 0.060&0.051&0.001& 3&  2\\
1105&13.999& 0.389&-0.035& 0.807& 2.766&0.024& 0.025& 0.045&0.043&0.009& 6&  3\\
1106&14.104& 0.399&-0.022& 0.816& 2.759&0.019& 0.023& 0.040&0.020&0.013& 6&  2\\
1108&13.929& 0.393&-0.023& 0.804& 2.764&0.026& 0.033& 0.064&0.029&0.013& 6&  3\\
\hline
\end{tabular}
\end{center}
 \label{tab:miembrosh3}
\end{table*}

\begin{table*}
\begin{center}
\caption{Photometric data for members of $h$ Persei (Continued Table 10).}
\begin{tabular}{crcrccccccccc}
\hline
Number&$V$&$b-y$&$m_{1}$&$c_{1}$&$\beta$&$\sigma_{V}$&$\sigma_{b-y}$&$\sigma_{m_{1}}$&$\sigma_{c_{1}}$&$\sigma_{\beta}$&$N_{uvby}$&$N_{\beta}$\\
\hline
1109&10.972& 0.340&-0.028& 0.198& 2.650&0.016& 0.020& 0.038&0.013&0.011& 5&  2\\
1110&13.461& 0.359&-0.001& 0.603& 2.739&0.021& 0.024& 0.056&0.048&0.013& 6&  3\\
1116& 9.252& 0.351&-0.050& 0.130& 2.617&0.014& 0.015& 0.025&0.021&  -  & 3&  1\\
1117&15.769& 0.518& 0.044& 0.993& 2.801&0.004& 0.005& 0.029&0.071&  -  & 2&  1\\
1118&14.085& 0.386&-0.031& 0.839& 2.767&0.022& 0.018& 0.029&0.023&0.018& 6&  3\\
1121&13.733& 0.388&-0.044& 0.683& 2.737&0.017& 0.021& 0.046&0.051&0.013& 6&  3\\
1122&12.224& 0.366& 0.004& 0.409& 2.714&0.018& 0.027& 0.056&0.050&0.009& 6&  3\\
1126&12.695& 0.344& 0.010& 0.318& 2.699&0.011& 0.027& 0.063&0.052&0.024& 6&  3\\
1128&12.153& 0.361&-0.012& 0.340& 2.675&0.011& 0.019& 0.049&0.046&0.016& 6&  3\\
1129&12.930& 0.391&-0.005& 0.507& 2.722&0.016& 0.032& 0.075&0.061&0.011& 6&  3\\
1130&14.474& 0.411&-0.032& 0.889& 2.850&0.035& 0.045& 0.070&0.064&  -  & 2&  1\\
1132& 8.440& 0.341&-0.017& 0.080& 2.588& -   &   -  &   -  &  -  &  -  & 1&  1\\
1133& 8.987& 0.367&-0.046& 0.089&  -   &0.002& 0.009& 0.018&0.014&  -  & 2&  0\\
1145&14.786& 0.424& 0.038& 1.025& 2.909&0.024& 0.037& 0.074&0.059&0.001& 4&  2\\
1147&14.611& 0.473& 0.010& 0.998& 2.861&0.026& 0.018& 0.027&0.054&0.041& 4&  2\\
1152&14.801& 0.432& 0.000& 1.058& 2.852&0.004& 0.011& 0.033&0.056&0.022& 4&  2\\
1163&15.132& 0.505&-0.006& 1.050& 2.849&0.011& 0.005& 0.039&0.070&0.001& 3&  2\\
1175&14.528& 0.427&-0.033& 1.105& 2.828&0.036& 0.027& 0.032&0.054&0.007& 5&  2\\
1177&14.441&0.443& 0.011& 0.824&2.806&0.039&0.031&0.073&0.062&0.031&5&2\\
1179&12.895& 0.386&-0.026& 0.464& 2.686&0.021& 0.030& 0.050&0.031&0.004& 6&  3\\
1180&14.829& 0.398& 0.157& 0.842& 2.915&0.002& 0.011& 0.012&0.064&0.023& 2&  2\\
1181&12.655& 0.352&-0.008& 0.338& 2.692&0.022& 0.022& 0.044&0.030&0.012& 6&  3\\
1185&13.178& 0.401&-0.002& 0.495& 2.744&0.016& 0.021& 0.063&0.074&0.011& 6&  3\\
1190&15.419& 0.458& 0.017& 1.134& 2.801&0.004& 0.009& 0.022&0.040&  -  & 3&  1\\
1191&14.713& 0.426& 0.023& 0.925& 2.879&0.015& 0.032& 0.078&0.045&0.019& 4&  2\\
1192&14.929& 0.454& 0.001& 1.236& 2.861&0.031& 0.060& 0.077&0.055&  -  & 2&  2\\
1198&13.804& 0.399&-0.027& 0.703& 2.715&0.025& 0.023& 0.039&0.086&0.009& 6&  3\\
1202&12.123& 0.352&-0.011& 0.356& 2.696&0.014& 0.026& 0.054&0.044&0.006& 6&  3\\
1203&13.910& 0.402&-0.022& 0.811& 2.788&0.030& 0.015& 0.035&0.130&0.048& 6&  3\\
1206&14.646& 0.435& 0.031& 0.988& 2.880&0.012& 0.031& 0.067&0.051&0.013& 4&  2\\
1213&14.622& 0.406&-0.007& 0.942& 2.866&0.022& 0.025& 0.061&0.053&0.039& 4&  2\\
1218&15.889& 0.530& 0.028& 1.028& 2.840&0.003& 0.010& 0.012&0.017&  -  & 2&  1\\
1222&15.639& 0.519& 0.028& 1.095& 2.917&0.002& 0.034& 0.071&0.090&  -  & 2&  2\\
1232&11.283& 0.338&-0.021& 0.202& 2.635&0.015& 0.020& 0.042&0.030&0.004& 6&  3\\
\hline
\end{tabular}
\end{center}
 \label{tab:miembrosh4}
\end{table*}

\begin{table*}
\begin{center}
\caption{Photometric data for members of $h$ Persei (Continued Table 10).}
\begin{tabular}{crcrccccccccc}
\hline
Number&$V$&$b-y$&$m_{1}$&$c_{1}$&$\beta$&$\sigma_{V}$&$\sigma_{b-y}$&$\sigma_{m_{1}}$&$\sigma_{c_{1}}$&$\sigma_{\beta}$&$N_{uvby}$&$N_{\beta}$\\
\hline
1240&15.905& 0.505& 0.066& 0.983& 2.834&  -  &  -   &  -   &  -  &  -  & 1&  1\\
1251&15.518& 0.486& 0.087& 1.122&  -   &0.036& 0.028& 0.037&0.050&  -  & 2&  0\\
1260&14.188& 0.434& 0.042& 0.814& 2.856&0.040& 0.028& 0.055&0.054&  -  & 4&  1\\
1262&14.423& 0.378&-0.007& 0.995& 2.862&0.067& 0.014& 0.012&0.062&  -  & 4&  1\\
1265&14.354& 0.355& 0.063& 0.902& 2.812&0.054& 0.025& 0.034&0.035&  -  & 4&  1\\
1267&14.422& 0.442& 0.041& 0.786& 2.848&0.025& 0.006& 0.001&0.007&  -  & 2&  1\\
1281&14.051& 0.425&-0.034& 0.927& 2.898&0.058& 0.009& 0.005&0.032&0.024& 2&  2\\
4005&15.569&0.599&0.003&0.962&2.776&0.023&0.043&0.082&0.035&0.023&3&2\\
4006&15.813&0.587&-0.027& 0.954&2.833&  -  &  -  &  -  &  -  & -   &1&1\\
4009&15.942& 0.687&-0.013& 0.864& 2.800&  -  &  -   &  -   &  -  &  -  & 1&  1\\
4011&15.976& 0.532& 0.093& 0.886& 2.826&0.004& 0.020& 0.087&0.086&  -  & 2&  1\\
4012&16.000& 0.704&-0.087& 0.827& 2.716&  -  &  -   &  -   &  -  &  -  & 1&  1\\
4013&16.023& 0.603&-0.050& 0.963& 2.793&  -  &  -   &  -   &  -  &  -  & 1&  1\\
4016&16.063& 0.623&-0.069& 0.784& 2.717&  -  &  -   &  -   &  -  &  -  & 1&  1\\
4017&16.097& 0.558&-0.044& 0.870& 2.790&  -  &  -   &  -   &  -  &  -  & 1&  1\\
4018&16.108& 0.627&-0.097& 1.022& 2.810&  -  &  -   &  -   &  -  &  -  & 1&  1\\
4023&16.158& 0.559& 0.025& 0.946& 2.779&  -  &  -   &  -   &  -  &  -  & 1&  1\\
4025&16.201& 0.565& 0.081& 0.710& 2.745&  -  &  -   &  -   &  -  &  -  & 1&  1\\
4029&16.330& 0.664&-0.108& 1.024& 2.840&  -  &  -   &  -   &  -  &  -  & 1&  1\\
4030&16.330& 0.648&-0.065& 0.781&  -   &  -  &  -   &  -   &  -  &  -  & 1&  0\\
4036&16.372& 0.603& 0.047& 0.718& 2.778&  -  &  -   &  -   &  -  &  -  & 1&  1\\
4037&16.385& 0.672&-0.126& 0.748& 2.764&  -  &  -   &  -   &  -  &  -  & 1&  1\\
4042&16.430& 0.643&-0.027& 0.821& 2.764&  -  &  -   &  -   &  -  &  -  & 1&  1\\
\hline
\end{tabular}
\end{center}
 \label{tab:miembrosh5}
\end{table*}

\begin{figure}
\begin{picture}(240,300)
\put(0,0){\includegraphics{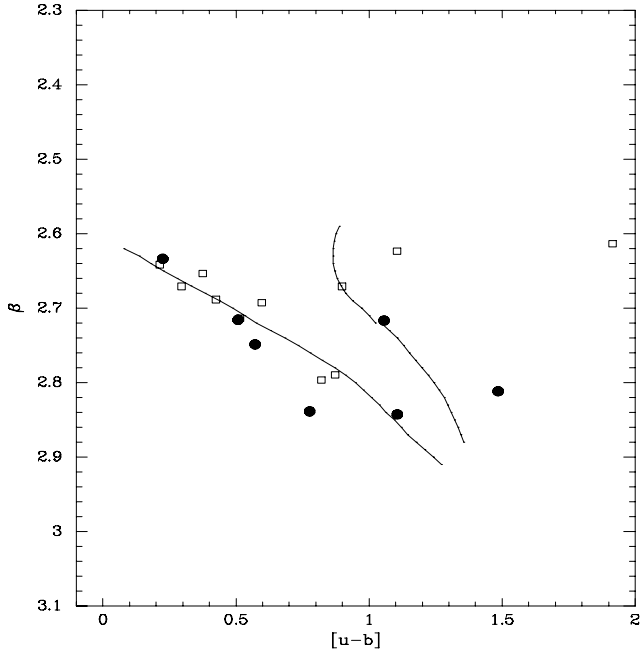}}
\end{picture}
\caption{[$u-b$] -- $\beta$ diagram for all stars whose position
deviates from the main sequence in only one of the photometric diagrams. 
The thin lines represent the loci of main-sequence B stars (left), 
the main sequence A-stars (bottom right) and the main sequence F-stars (top 
right).
Open squares represent stars from the field 
of $h$ Persei and while filled circles are from the field of $\chi$ Persei.}
\label{fig:ubbeta}
\end{figure}

There is a smaller number of stars whose position is displaced with
respect to the main sequence
in only one of these diagrams.
For these stars we calculate the free reddening indices [$m_{1}$],
[$c_{1}$] and [$u-b$]: 

\begin{equation}
[m_{1}]=m_{1}+0.32(b-y)
\end{equation}
\begin{equation}
[c_{1}]=c_{1}-0.20(b-y)
\end{equation}
\begin{equation}
[u-b]=[c_{1}]+2[m_{1}]
\end{equation}

and use the $\beta - [u-b]$ diagram to roughly evaluate their spectral
classification  (See Figure~\ref{fig:ubbeta}).
We find that stars 867, 969, 1084 
and 1138 in $h$ Persei and stars 2376, 7029 and 7081 in $\chi$ Persei
have spectral types not corresponding to
their photometric positions. 
We consider these objects as non-members.
As the $c_{1}$ index is less affected by extinction than the $(b-y)$ 
colour, the $V-c_{1}$ diagram provides a more secure diagnostic than
the $V-(b-y)$ diagram.
Therefore, stars whose position only deviates from
the main sequence in the  $V-(b-y)$ plane could be affected by differential
extinction. In $\chi$ Persei, the only case is star 2370. Since Muminov (1980)
gives it a relatively low membership probability ($0.38$), we have excluded 
this star from further analysis.
In $h$ Persei, we have stars $911$, $1072$ and $1257$. The membership 
probability is low for 911 and 1257 (0.15 and 0.36 respectively
according to Muminov 1980). So we consider them as likely non-members of the 
cluster. Star 1072, on the other hand, has much higher probability, 0.78 
(Muminov 1980), and we consider it a cluster member.
Other stars, 2185, 2140 and 2311 in $\chi$ Persei have positions in 
the $V-c_{1}$ diagram deviating sligthly towards brighter $V$. 
The membership
probability is high for 2185 and 2140 (0.76 and 0.98 respectively in Muminov 
1980). We consider 2185 and 2140 as
members. The case for 2311 is not so clear, but since it is likely to be an
eclipsing
binary (Krzesi\'{n}ski \& Pigulski 1997), we will not use it for the determination of 
the cluster parameters, though 
it may well be a member (Vrancken et al. 2000). A similar situation occurs with the 
stars 859, 869, 926 and 1000. 
In $h$ Persei we consider as non-member the star 859, because of its low membership 
probability (0.07; Muminov 1980), and as members stars 869, 926 and 1000 
because of their higher probabilities ($0.74$, $0.60$ and $0.99$ 
respectively; Muminov 1980).         
In Tables 10, 11, 12, 13, 14, 15, 16, 17, 18 and 19 we present the resulting 
values for $V, (b-y), m_{1}, c_{1}$ and $\beta$ and their precisions,
together with the number of observations for member stars of $h$ \&
$\chi$ Persei respectively.
In Table 20 and Table 21, we list the 
stars identified as non-members in the fields of $h$ \& $\chi$ Persei 
respectively together with their photometric data.
Finally in Table 22 and Table 23, 
we give the photometric data for Be star members of $h$ \& $\chi$ Persei 
respectively.   
From the list of member stars we select those falling in the range of B-type stars in the V-(b-y) and V-$c_{1}$ diagrams.
From this subset we remove all 
Be stars and all early B stars which deviate strongly from the ZAMS. We also exclude stars 869, 926, 976, 1000, 1072, 2140 and 
2185 because, in spite of appearing as likely members in the photometric 
diagrams,
they have positions deviating slightly from the average loci of stars 
with the same spectral type.
For all the remaining B stars we calculate individual reddenings .We
follow the procedure described by Crawford et al. (1970b): we use the 
observed $c_{1}$ to
predict the first approximation to $(b-y)_{0}$ with the expresion $(b-y)_{0}=-0.116+0.097c_{1}$. Then we calculate $E(b-y)=(b-y)-(b-y)_{0}$ and use
$E(c_{1})=0.2E(b-y)$ to correct $c_{1}$ for reddening $c_{0}=c_{1}-E(c_{1})$.
 The intrinsic color $(b-y)_{0}$ is now calculated by replacing $c_{1}$ by 
$c_{0}$ in the above equation for $(b-y)_{0}$. Three iterations are 
enough to reach convergence in the
process.  
The final average values of reddening are $E(b-y)= 0.44\pm0.02$ for
$h$ Persei and $E(b-y)=0.39\pm0.05$ for $\chi$ Persei. These values are consistent,
within the errors,  with both clusters having
the same reddening in their central region.   

\begin{figure}
\begin{picture}(240,300)
\put(0,0){\includegraphics{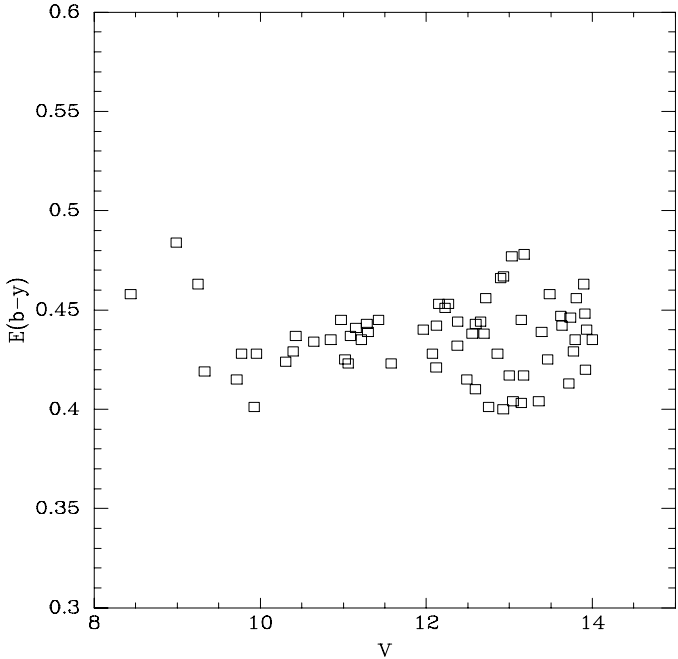}}
\end{picture}
\caption{Individual values of $E(b-y)$ calculated using Crawford's et al. (1970b)
procedure against $V$ magnitude for $h$ Persei members in the B spectral type
range.}
\label{fig:exceso1}
\end{figure}

\begin{figure}
\begin{picture}(240,300)
\put(0,0){\includegraphics{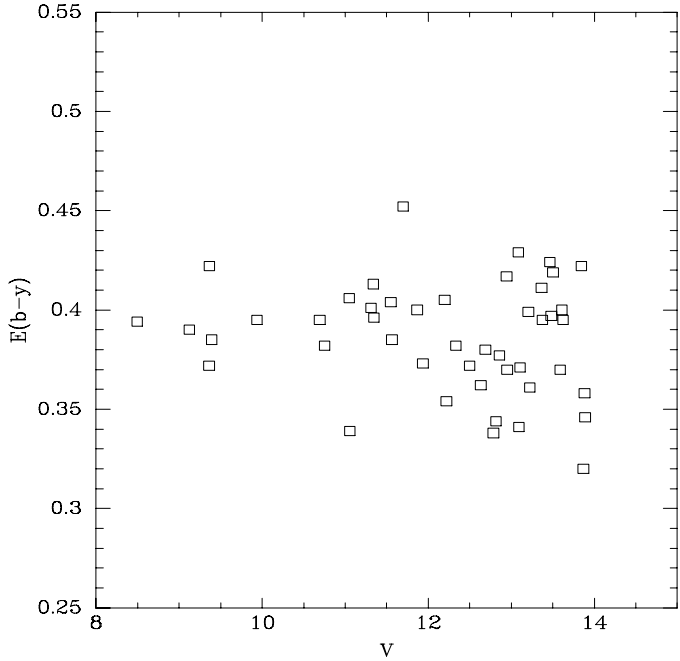}}
\end{picture}
\caption{Individual values of $E(b-y)$ calculated using Crawford's et al. (1970b)
procedure against $V$ magnitude for $\chi$ Persei members in the B spectral type
range.}
\label{fig:exceso2}
\end{figure}

In the $E(b-y)$ -- $V$ diagrams (see Figure~\ref{fig:exceso1} and 
Figure~\ref{fig:exceso2}) we notice that the scatter of individual reddenings is 
quite small in both clusters, confirming the result obtained by 
Crawford et al. (1970b) of homogeneous reddening across the central region.  
With the aid of these values we calculate the intrinsic colours and magnitudes of the candidate
members of both clusters, listed in Tables 24, 25, 26 and 27. 
\begin{table*}
\begin{center}
\caption{Photometric data and spectral type for bright members of 
$h$ \& $\chi$ Persei taken from the literature. The fifth column
indicates whether the star is inside the area covered by our observations (In)
or more distant from the central region (Out).}
\begin{tabular}{lrccccc}
\hline
\multicolumn{7}{c}{$h$ Persei}\\
\hline
$Number$&$V$&$(b-y)$&$c_{1}$&$E(b-y)$&$Position$&$Spectral Type$\\
\hline
3&7.400&0.244&0.051&0.360&Out&B2Ib\\
339&8.850&0.288&0.071&0.410&Out&B1IV\\
612&8.410&0.255&0.070&0.380&Out&B1II\\
1162&6.660&0.443&0.101&0.560&In&B2Ia\\
1187&10.820&0.348&0.212&0.460&In&B2IV\\
1899&8.530&0.289&0.138&0.400&Out&B2II\\
1781&9.210&0.267&0.145&0.380&Out&B1IV\\
\hline
\multicolumn{7}{c}{$\chi$ Persei}\\
\hline
2227&8.050&0.328&0.111&0.440&In&B2II\\
2361&8.750&0.351&0.108&0.460&Out&B0.5III\\
2541&9.150&0.292&0.146&0.400&Out&B2II\\
2589&7.440&0.574&0.860&0.610&Out&A2Iap?\\
2621&7.000&0.512&0.455&0.590&Out&B8Ia\\
\hline
\end{tabular}
\end{center}
 \label{tab:liter}
\end{table*}   

Since none of the stars in the sample,
except RS Per, is in a late evolutionary state
(all stars earlier than B3 deviate from
the main sequence, but none seems to have a luminosity class higher
than III), we have added to our colour-magnitude plots all the stars
brighter than $V=11$ taken from Johnson \& Morgan (1955) and Crawford et
al. (1970b), which, as shown in Section~\ref{sec:rdctn}, are on the
same system as our observations. Values of $V$ (Johnson \& Morgan
1955), $(b-y)$, $c_{1}$, individual $E(b-y)$, position in the clusters
(Crawford et al. 1970b) and spectral type (Schild 1965; Slettebak 1968)
are given in Table 15. A few of these stars are inside the area covered
by our observations but were saturated in some of our frames. 

As we can see from the values of individual reddening, the $h$ Persei
stars outside the inner $5.6$ $arcmin$ have reddenings
lower that the average calculated for the inner $5.6$ $arcmin$, and
therefore they lie to the left of the rest of the members
in the $V-(b-y)$ diagram. This is due to the fact that the $(b-y)$ 
colour is more affected by
reddening than the $c_{1}$ index. Their position in the $V-c_{1}$ diagram
agrees well with the rest of the cluster, because this index is
very little affected by reddening. 

In $\chi$ Persei three stars
outside the central region have reddenings higher than the average for the
central region, and therefore they lie to the right of the rest 
of the stars in the $V-(b-y)$ plane.
Once more their position in the $V-c_{1}$ diagram
is compatible with the rest of the cluster. We conclude that the
analysis of the $V-c_{1}$ diagram yields much firmer results than the 
analysis of the $V-(b-y)$ diagram, as the
actual position of each star in the latter is modified by the
difference between its individual reddening and the average value for the
central region.

\subsection{Spectral Classification}

Given the
paucity of spectroscopic studies of cluster members, we decided to estimate
the spectral types for all members observed by us. Since our objects cover
a wide
range of spectral types, we use different procedures depending on the
intrinsic photometric values. Following Napiwotzki et al. (1993), we 
select stars with $(b-y)_{0} \leq 0.00$
 (corresponding to $T_{\rm eff} \ga 9500\:$K) and use
the temperature calibration based on the dereddened [$u-b$] index, given by  
Napiwotzki et al. (1993):
\begin{equation}
\Theta \equiv \frac{5040\:{\rm K}}{T_{\rm eff}} = 0.1692 + 0.2828[u-b] -
[u-b]^{2}
\end{equation}
Once the temperature of the star is known, we 
derive an approximate value for the gravity
$\log g$ by means of the grids of Moon \& Dworetsky (1985). For the 
hottest
stars ($T_{\rm eff} \ga 20000\:{\rm K}$), only a rough
estimate of $\log g$ is derived from the theoretical grids of 
Lester et al. (1986).

For stars with $0.00 \leq (b-y)_{0} \leq 0.04$ (corresponding to 
$8500\:{\rm K} \la T_{\rm eff} \la 9500\:$K), we use the parameters
\begin{equation}
a_{0} = 1.36(b-y)_{0} + 0.36m_{0} + 0.18 c_{0} - 0.2448
\end{equation}
and
\begin{equation}
r^{*}=0.35c_{1} - 0.07(b-y) -(\beta -2.565)
\end{equation}
to derive both $T_{\rm eff}$ and $\log g$ from the grids of Moon \& 
Dworetsky (1985).

\begin{table*}
\begin{center}
\caption{Photometric data for members of $\chi$ Persei.}
\begin{tabular}{crcrccccccccc}
\hline
Number&$V$&$b-y$&$m_{1}$&$c_{1}$&$\beta$&$\sigma_{V}$&$\sigma_{b-y}$&$\sigma_{m_{1}}$&$\sigma_{c_{1}}$&$\sigma_{\beta}$&$N_{uvby}$&$N_{\beta}$\\
\hline
2085&11.310&0.294&-0.039&0.276&2.664&0.020&0.010&0.002&0.027&0.023&2&2\\
2091&11.699&0.361&-0.082&0.348&2.585&0.019&0.022&0.023&0.008&0.010&3&3\\
2092&14.763&0.417&-0.069&1.082&2.864&0.026&0.020&0.040&0.024&0.043&3&2\\
2094&11.866&0.302&-0.077&0.262&2.659&0.017&0.027&0.051&0.048&0.011&6&4\\
2108&14.613&0.344&0.047 &0.950&  -  &  -  &  -  & -   &  -  & -   &1&0\\
2109&15.017&0.449&-0.070&1.083&2.893&0.005&0.005&0.035&0.058&0.013&3&2\\
2111&13.464&0.366&-0.085&0.686&2.739&0.017&0.012&0.024&0.027&0.002&6&4\\
2114&11.046&0.308&-0.101&0.262&2.641&0.011&0.017&0.032&0.021&0.005&5&3\\
2116&13.503&0.362&-0.059&0.696&2.745&0.017&0.019&0.031&0.049&0.016&6&4\\
2123&15.053&0.411&-0.031&1.122&2.893&0.011&0.028&0.062&0.013&0.009&3&2\\
2124&15.139&0.464&-0.038&1.097&2.889&0.021&0.011&0.016&0.021&0.025&3&2\\
2133&12.198&0.317&-0.087&0.371&2.678&0.013&0.013&0.018&0.015&0.013&6&4\\
2139&11.338&0.316&-0.136&0.279&2.636&0.007&0.009&0.033&0.042&0.006&5&3\\
2147&14.359&0.392&-0.083&1.022&2.802&0.023&0.034&0.051&0.051&0.027&5&2\\
2149&14.133&0.385&-0.056&0.838&2.785&0.021&0.022&0.039&0.044&0.050&6&3\\
2155&15.630&0.458&-0.085&0.931&2.843&0.018&0.020&0.059&0.102&-&3&1\\
2167&13.364&0.347&-0.080&0.616&2.740&0.012&0.008&0.012&0.024&0.007&6&4\\
2170&15.298&0.454&-0.042&1.109&2.921&0.019&0.042&0.071&0.035&0.013&3&2\\
2174&15.315&0.437&-0.008&1.114&2.854&0.015&0.025&0.044&0.036&0.022&3&2\\
2175&14.555&0.360&-0.015&1.030&2.846&0.003&0.006&0.011&0.046& -   &3&1\\
2179&14.139&0.341&0.011 &0.988&2.812&0.022&0.036&0.086&0.076&0.011&6&4\\
2185&10.927&0.270&-0.005&0.399&2.716&0.007&0.014&0.039&0.045&0.025&4&3\\
2193&15.402&0.439&0.048 &1.048&2.866&0.018&0.020&0.036&0.019& -   &2&1\\
2194&13.481&0.340&-0.060&0.686&2.741&0.020&0.027&0.043&0.051&0.030&6&4\\
2196&11.549&0.304&-0.066&0.246&2.646&0.012&0.013&0.023&0.023&0.010&6&4\\
2200&12.688&0.303&-0.041&0.475&2.709&0.010&0.017&0.034&0.038&0.008&6&4\\
2203&14.657&0.370&-0.009&1.058&2.858&0.009&0.006&0.014&0.022&0.015&3&2\\
2206&15.590&0.482&-0.003&1.014&2.849&0.040&0.030&0.040&0.095& -   &3&1\\
2209&15.396&0.424&0.062 &1.141&2.864&0.002&0.025&0.031&0.040& -   &2&1\\
2211&12.950&0.300&-0.009&0.552&2.705&0.024&0.034&0.064&0.054&0.021&6&4\\
2214&15.774&0.471&0.029 &0.976&2.862&0.051&0.004&0.007&0.024& -   &2&1\\
2215&15.546&0.482&-0.026&1.112&2.894&0.030&0.008&0.023&0.098& -   &2&1\\
2219&15.423&0.501&0.022 &0.948&2.831&0.017&0.029&0.031&0.073&0.050&3&2\\
2223&14.018&0.391&-0.021&1.052&2.881&0.026&0.028&0.042&0.041&0.028&6&4\\
2224&13.846&0.366&-0.053&0.702&2.761&0.022&0.019&0.044&0.043&0.025&6&4\\
2229&11.343&0.295&-0.067&0.238&2.642&0.002&0.011&0.017&0.006&0.012&5&3\\
\hline
\end{tabular}
\end{center}
 \label{tab:miembroschi1}
\end{table*}

\begin{table*}
\begin{center}
\caption{Photometric data for members of $\chi$ Persei (Continued Table 16).}
\begin{tabular}{crcrccccccccc}
\hline
Number&$V$&$b-y$&$m_{1}$&$c_{1}$&$\beta$&$\sigma_{V}$&$\sigma_{b-y}$&$\sigma_{m_{1}}$&$\sigma_{c_{1}}$&$\sigma_{\beta}$&$N_{uvby}$&$N_{\beta}$\\
\hline
2232&11.054&0.233&-0.018&0.168&2.640&0.012&0.016&0.041&0.036&0.019&5&3\\
2235& 9.365&0.311&-0.071&0.131&2.590&0.012&0.015&0.046&0.039&0.022&3&2\\
2239&14.220&0.367&-0.036&0.854&2.797&0.022&0.048&0.073&0.022&0.026&3&3\\
2240&14.036&0.285&0.076 &0.759&2.821&0.038&0.034&0.082&0.089&0.015&6&4\\
2241&13.590&0.319&-0.012&0.747&2.779&0.026&0.028&0.053&0.047&0.031&6&4\\
2245&12.497&0.289&-0.035&0.419&2.671&0.016&0.016&0.028&0.016&0.005&6&4\\
2246& 9.936&0.282&-0.043&0.112&2.599&0.011&0.027&0.060&0.044&0.023&4&3\\
2249&15.149&0.381&0.042 &1.080&2.938&0.015&0.011&0.039&0.041&0.020&3&2\\
2251&11.563&0.297&-0.028&0.367&2.703&0.009&0.011&0.027&0.030&0.012&6&4\\
2253&12.633&0.288&0.007 &0.505&2.732&0.011&0.018&0.052&0.048&0.028&6&4\\
2254&15.272&0.447&-0.001&1.061&2.916&0.010&0.025&0.039&0.026&0.001&3&2\\
2255&10.692&0.288&-0.047&0.174&2.607&0.009&0.018&0.035&0.021&0.008&5&3\\
2258&13.948&0.329&-0.019&0.882&2.807&0.022&0.027&0.046&0.069&0.041&6&3\\
2260&13.977&0.376&-0.030&0.847&2.822&0.025&0.042&0.071&0.054&0.062&6&3\\
2261&14.534&0.339&0.003 &1.008&2.865&0.017&0.028&0.065&0.065&0.013&4&1\\
2267&13.224&0.287&0.021 &0.507&2.747&0.014&0.020&0.061&0.056&0.014&6&4\\
2268&13.622&0.336&-0.030&0.669&2.741&0.014&0.021&0.039&0.045&0.035&6&4\\
2269&13.104&0.302&-0.007&0.558&2.727&0.008&0.018&0.059&0.077&0.018&6&4\\
2270&14.153&0.378&-0.023&0.925&2.785&0.028&0.018&0.039&0.047&0.029&6&4\\
2275&13.086&0.352&-0.048&0.489&2.719&0.016&0.015&0.029&0.018&0.037&6&4\\
2277&14.948&0.368&0.066 &1.077&2.936&0.017&0.019&0.051&0.056&0.034&3&2\\
2283&14.640&0.372&-0.010&0.991&2.855&0.011&0.014&0.036&0.023&0.035&3&2\\
2286&14.725&0.376&0.013 &1.053&2.887&0.015&0.035&0.066&0.088&0.034&3&2\\
2294&14.158&0.322&0.012 &0.811&2.835&0.025&0.016&0.054&0.056&0.028&5&4\\
2296& 8.493&0.280&-0.023&0.096&2.591&0.004&0.019&0.065&0.080&0.019&3&2\\
2297&12.943&0.339&-0.058&0.476&2.699&0.018&0.020&0.028&0.030&0.006&6&4\\
2299& 9.124&0.272&-0.017&0.061&2.594&0.019&0.016&0.055&0.049&0.007&3&2\\
2300&15.598&0.452&0.058 &0.742&  -  & -   &  -  & -   &  -  &  -  &1&0\\
2301&11.934&0.282&0.002 &0.336&2.693&0.023&0.016&0.046&0.040&0.020&6&4\\
2307&15.195&0.424&0.049 &1.046&2.836&0.030&0.032&0.079&0.077&0.015&3&2\\
2309&12.784&0.262&0.028 &0.481&2.703&0.016&0.029&0.078&0.066&0.007&6&3\\
2311& 9.390&0.284&-0.065&0.230&2.634& -   & -   & -   &  -  &  -  &1&1\\
2314&14.560&0.351&0.026 &1.072&2.885&0.011&0.022&0.069&0.026&0.016&3&2\\
2316&14.482& 0.356& -0.009&0.985&2.837&0.020&0.033&0.045&0.034&0.012&3&2\\
2317&15.005&0.388&0.043 &1.089&2.891&0.031&0.033&0.082&0.055&0.028&3&2\\
2319&13.092&0.272&0.026 &0.549&2.725&0.020&0.030&0.083&0.071&0.019&6&4\\
\hline
\end{tabular}
\end{center}
 \label{tab:miembroschi2}
\end{table*}

\begin{table*}
\begin{center}
\caption{Photometric data for members of $\chi$ Persei (Continued Table 16).}
\begin{tabular}{crcrccccccccc}
\hline
Number&$V$&$b-y$&$m_{1}$&$c_{1}$&$\beta$&$\sigma_{V}$&$\sigma_{b-y}$&$\sigma_{m_{1}}$&$\sigma_{c_{1}}$&$\sigma_{\beta}$&$N_{uvby}$&$N_{\beta}$\\
\hline
2323&14.218&0.340&0.017 &0.920&2.830&0.024&0.035&0.078&0.085&0.015&4&3\\
2324&13.889&0.303&0.024 &0.822&2.768&0.030&0.036&0.071&0.052&0.007&5&4\\
2331&14.185&0.327&0.031 &0.951&2.852&0.028&0.029&0.045&0.043&0.022&6&4\\
2332&14.578&0.421&-0.083&0.988&2.875&0.023&0.046&0.087&0.067&0.001&3&2\\
2335&14.406&0.405&-0.003&1.046&2.847&0.045&0.063&0.081&0.027&0.017&4&2\\
2338&13.610&0.345&-0.016&0.711&2.790&0.019&0.017&0.040&0.025&0.015&6&4\\
2349&12.815&0.264&0.051 &0.440&2.706&0.015&0.040&0.101&0.072&0.006&6&4\\
2350&13.376&0.322&-0.004&0.525&2.717&0.024&0.017&0.040&0.023&0.007&6&4\\
2352&12.332&0.298&0.005 &0.404&2.700&0.020&0.025&0.052&0.024&0.015&6&4\\
2355&15.574&0.430&0.139 &0.877&2.919&0.056&0.045&0.135&0.208& -   &2&1\\
2358&14.308&0.319&0.037 &0.859&2.845&0.017&0.027&0.071&0.061&0.003&5&2\\
2359&13.882&0.312&0.015 &0.791&2.812&0.034&0.025&0.062&0.071&0.013&6&4\\
2362&15.446&0.415&0.054 &1.066&2.949&0.013&0.026&0.040&0.027&0.019&3&2\\
2363&13.920&0.307&0.039 &0.851&2.841&0.022&0.031&0.079&0.050&0.011&6&4\\
2379&12.218&0.261&0.023 &0.310&2.683&0.023&0.036&0.090&0.062&0.008&6&4\\
2392&10.752&0.276&-0.041&0.177&2.653&0.017&0.020&0.018&0.004&0.016&2&2\\
2401&15.458&0.538&-0.008&1.065&2.930&0.038&0.022&0.010&0.110& -   &2&1\\
2407&14.324&0.326&0.073 &0.839&  -  &0.038&0.015&0.015&0.055&  -  &2&0\\
2410&14.210&0.326&0.059 &0.787&  -  &0.033&0.020&0.055&0.044&  -  &3&0\\
2414&13.869&0.284&0.136 &0.890&2.811&0.022&0.050&0.110&0.070&  -  &3&1\\
2416&15.313&0.376&0.152 &0.988&  -  &  -  &  -  &  -  &  -  &  -  &1&0\\
2417& 8.956& 1.733&-0.301&1.181&2.703&0.018& 0.006&0.009&0.102&0.000&2&2\\
7014&15.887&0.558&-0.013&0.893&2.785&0.022&0.005&0.042&0.052&  -  &2&1\\
7015&15.930&0.544&-0.027&0.911&2.842&  -  &  -  &  -  &  -  &  -  &1&1\\
7016&15.703&0.496&0.023 &1.003&2.880&0.004&0.011&0.017&0.018&  -  &2&1\\
7017&16.163&0.596&-0.008&0.782&2.696&  -  &  -  &  -  &  -  & -   &1&1\\
7021&16.271&0.664&-0.094&0.786&2.694&  -  &  -  &  -  &  -  &  -  &1&1\\
7023&16.060&0.624&-0.074&0.833&2.793&  -  &  -  &  -  &  -  &  -  &1&1\\
7024&16.006&0.574&-0.031&0.850&2.788&0.040&0.003&0.007&0.068& -   &2&1\\
7027&15.713&0.575&-0.039&0.941&2.780&0.015&0.044&0.059&0.061& -   &2&1\\
7037&16.295&0.630&-0.054&0.643&  -  &  -  &  -  &  -  &  -  & -   &1&0\\
7038&15.143&0.414&-0.024&1.143&2.939&0.036&0.040&0.061&0.002&0.025&2&2\\
7045&15.652&0.556&0.048 &0.754&2.748&0.009&0.003&0.004&0.062& -   &2&1\\
7046&15.937&0.455&-0.007&1.119&2.790&0.030&0.002&0.003&0.039& -   &2&1\\
7047&16.102&0.572&-0.057&0.899&2.845&  -  &  -  &  -  &  -  & -   &1&1\\
7048&15.735&0.510&-0.044&1.072&2.891&0.023&0.023&0.020&0.007&  -  &2&1\\
\hline
\end{tabular}
\end{center}
 \label{tab:miembroschi3}
\end{table*}

\begin{table*}
\begin{center}
\caption{Photometric data for members of $\chi$ Persei (Continued Table 16).}
\begin{tabular}{crcrccccccccc}
\hline
Number&$V$&$b-y$&$m_{1}$&$c_{1}$&$\beta$&$\sigma_{V}$&$\sigma_{b-y}$&$\sigma_{m_{1}}$&$\sigma_{c_{1}}$&$\sigma_{\beta}$&$N_{uvby}$&$N_{\beta}$\\
\hline
7049&16.481&0.656&-0.146&0.699&0.000&   - &  -  &   - &  -  &  -  &1&0\\
7052&16.063&0.579&0.011 &0.724&2.852&   - &   - &  -  & -   & -   &1&1\\
7054&16.069&0.588&-0.032&0.779&2.718&   - &  -  & -   &   - & -   &1&1\\
7062&15.102&0.398&0.018 &1.037&2.905&0.018&0.026&0.049&0.041&0.002&3&2\\
7064&15.932&0.517&0.025 &0.974&2.784&0.010&0.011&0.014&0.024& -   &2&1\\
7066&15.925&0.483&0.030 &0.912&2.972&  -  & -   &  -  &   - &  -  &1&1\\
7067&16.274&0.602&-0.031&0.788&2.678&  -  & -   &  -  &   - &  -  &1&1\\
7068&15.910&0.528&0.013 &0.918&2.802&0.012&0.010&0.000&0.105&     &2&1\\
7070&16.118&0.608&-0.080&0.802&2.778&  -  &  -  &  -  &  -  &  -  &1&1\\
7077&16.098&0.579&-0.071&0.899&2.786&  -  &  -  &  -  &  -  &  -  &1&1\\
7080&15.859&0.587&-0.079&0.950&2.842&  -  &  -  &  -  &  -  &  -  &1&1\\
7083&15.566&0.443&0.061 &0.987&2.852&0.048&0.022&0.069&0.032&  -  &2&1\\
7084&13.206&0.327&-0.010&0.530&  -  &0.009&0.028&0.063&0.050&  -  &6&0\\
7085&15.874&0.496&0.092 &0.760&2.832&0.056&0.036&0.101&0.080&  -  &2&1\\
7086& 9.361&0.256&0.036 &0.074&2.615&0.025&0.008&0.004&0.006&0.005&2&2\\
7088&16.500&0.719&-0.093&0.581&  -  &  -  &  -  &  -  &  -  & -   &1&0\\
7091&16.212&0.707&-0.146&0.682&2.719&  -  &  -  &  -  &  -  &  -  &1&1\\
7092&14.839&0.379&0.058 &1.040&2.891&0.011&0.020&0.041&0.006&0.028&3&2\\
7093&16.014&0.541&0.042 &0.764&2.762&0.041&0.008&0.037&0.004&  -  &2&1\\
7096&15.987&0.569&-0.021&0.780&2.820&  -  &  -  &  -  &   - &   - &1&1\\
7097&16.119&0.719&-0.082&0.627&2.677&  -  &  -  &  -  &  -  &   - &1&1\\
7099&16.105&0.600&-0.021&0.807&2.712&  -  &  -  &  -  &  -  &   - &1&1\\
7104&15.763&0.473&0.078 &0.939&2.835&0.009&0.010&0.066&0.034&  -  &2&1\\
7105&16.083&0.562&-0.087&0.974&2.782&  -  &  -  &  -  &  -  &  -  &1&1\\
7108&16.443&0.641&-0.034&0.632&2.705&  -  &  -  &  -  &  -  &  -  &1&1\\
7109&16.085&0.564&-0.009&0.875&2.767&  -  &  -  &  -  &  -  &  -  &1&1\\
7116&16.011&0.597&-0.058&0.859&2.741&  -  &  -  &  -  &  -  &  -  &1&1\\
7118&15.893&0.459&0.085 &0.898&2.799&0.040&0.007&0.041&0.021&   - &2&1\\
7122&14.881&0.427&0.083 &0.899&2.900&0.018&0.061&0.137&0.085&0.017&3&2\\
\hline
\end{tabular}
\end{center}
 \label{tab:miembroschi4}
\end{table*}

Finally, for stars with $(b-y)_{0} \geq 0.04$, we derive $T_{\rm eff}$ 
and $\log g$ directly from $c_{0}$ and $\beta$ by using the grids of 
Moon \& Dworetsky (1985).

As a last step, we derive approximate spectral types by
correlating the estimated $T_{\rm eff}$ and $\log g$ with the average
values for each spectral type from Kontizas \& Theodossiou (1980) and
Allen (1973). The estimated spectral types are listed in 
Tables 24, 25, 26 and 27. The validity of this approximation is confirmed
by the fact that, for all the stars which have a spectroscopic spectral 
classification, our estimate is consistent with the spectroscopic determination
with an uncertainty of $\pm1$ subtype. The only exception is 
the star
1116, for which we give a spectral type B2, while a spectral type B0.5V
is given by Schild (1965). We note that, based on
Geneva photometry, Waelkens et
al. (1990) found that none of the stars classified as B0.5 in $h$ Persei
seems to be any hotter than other stars classified as B1 or B1.5, showing
that the spectral classification of some stars is uncertain.

\begin{table*}
\begin{center}
\caption{Photometric data for non-members of $h$ Persei.}
\begin{tabular}{crcrccccccccc}
\hline
Number&$V$&$b-y$&$m_{1}$&$c_{1}$&$\beta$&$\sigma_{V}$&$\sigma_{b-y}$&$\sigma_{m_{1}}$&$\sigma_{c_{1}}$&$\sigma_{\beta}$&$N_{uvby}$&$N_{\beta}$\\
\hline
 859&10.726&0.352&-0.126& 0.311&2.642&0.018&0.015&0.024&0.021&0.004&5&2\\
 865&13.276&0.663& 0.538& 0.283&2.577&0.015&0.012&0.021&0.023&0.013&4&3\\
 867&10.572&0.376& 0.177& 0.379&2.671&0.006&0.008&0.013&0.025&0.007&5&2\\
 886&15.346&0.749&-0.054& 0.644&2.614&0.009&0.019&0.031&0.033&0.006&3&2\\
 911&11.359&0.287&-0.033& 0.235&2.671&0.013&0.010&0.015&0.009&0.010&4&2\\
 969&13.017&0.945& 0.100& 0.490&2.624&0.008&0.017&0.039&0.016&0.019&5&3\\
1015&10.573&0.222& 0.030& 0.677&-    &0.037&0.023&0.006&0.031&-    &2&0\\
1023&12.567&0.413& 0.197& 0.266&2.659&0.021&0.016&0.030&0.018&0.005&5&3\\
1047&12.039&0.268& 0.062& 0.168&2.705&0.012&0.010&0.029&0.015&0.004&5&3\\
1054&13.831&0.623& 0.111& 0.506&2.606&0.017&0.010&0.021&0.045&0.026&6&3\\
1084&15.940&0.510& 0.022& 0.552&2.797&  -  &  -  &  -  &  -  & -   &1&1\\
1099&13.179&0.436&-0.042& 0.957&2.834&0.014&0.017&0.022&0.031&0.005&6&3\\
1100&14.060&0.562& 0.094& 0.677&2.761&0.030&0.041&0.102&0.084&0.018&6&3\\
1138&12.710&1.222& 0.439& 0.500&2.614&0.015&0.021&0.055&0.049&0.031&3&2\\
1155&12.518&0.400&-0.009& 0.616&2.769&0.016&0.027&0.055&0.032&0.023&6&3\\
1167&14.028&0.571& 0.129& 0.398&2.628&0.032&0.020&0.054&0.044&0.013&6&3\\
1184&12.113&0.317& 0.061& 0.183&2.712&0.013&0.031&0.062&0.047&0.008&4&2\\
1189&13.810&0.639& 0.147& 0.579&2.615&0.016&0.012&0.069&0.165&0.027&4&2\\
1194&14.598&1.025& 0.233& 0.398&2.618&0.025&0.017&0.068&0.075&0.013&2&2\\
1238&15.324&0.743& 0.003& 0.510&2.656&  -  &  -  &  -  &  -  & -   &1&1\\
1257&10.326&0.403& 0.029& 0.139&2.654&0.020&0.018&0.029&0.021&0.004&2&2\\
1272&13.406&1.007& 0.179& 0.457&2.604&0.016&0.023&0.030&0.015&0.028&2&2\\
4001&15.302&0.722&-0.001& 0.512&2.650&0.012&0.003&0.011&0.054& -   &3&1\\
4002&15.446&1.006& 0.049& 0.333&2.623&  -  &  -  &  -  &  -  & -   &1&1\\
4007&15.838&0.721&-0.081& 0.598&2.689&  -  &  -  &  -  &  -  & -   &1&1\\
4014&16.044&0.701&-0.002& 0.483&2.610&  -  &  -  &  -  &  -  & -   &1&1\\
4019&16.135&0.801&-0.122& 0.527&2.585&  -  &  -  &  -  &  -  & -   &1&1\\
4021&16.162&0.782&-0.033& 0.393&2.615&  -  &  -  &  -  &  -  & -   &1&1\\
4040&16.400&0.459&-0.107& 1.140&2.818&0.007&0.030&0.007&0.005& -   &2&1\\
\hline
\end{tabular}
\end{center}
 \label{tab:nomiembrosh}
\end{table*} 

For the vast majority of the stars in the sample, we derive gravities
compatible with their being main-sequence objects. Among the F and late
A stars the scatter in the derived gravities is rather larger than among
B-type stars, probably reflecting the larger errors associated with fainter
magnitudes. For almost all of the stars at the low temperature end, we 
derive low gravities ($\log g \la 3.6$). Among the stars later
than F0, only 7097 (F4), 7108 (F2) and 7091 (F2) give $\log g \ga 3.7$.
The situation completely reverses for the hot stars. On the whole A0-F0
range, only star 856 (A7) has a value of $\log g$ that stands out as 
being particularly lower than that of all other stars. Among B stars, where
the gravity determination is probably more reliable, a few stars have
$\log g \approx 3.5$, and could be evolved. These are 1198 (B9), 1020 (B8),
1179 (B6) and 1232 (B3). From their atmospheric parameter calculations,
Vrancken et al. (2000) find that B1 and B1.5 stars classified as 
main-sequence have gravities corresponding to higher luminosities.
In particular, they find $\log g = 3.4$ for the star 2311 (B2III), for which
we obtain  $\log g \approx 3.5$. This would imply that also the star 2255 (B2)
and the star 2246 (B1), for which we estimate a similar gravity, are giants.
\begin{table*}
\begin{center}
\caption{Photometric data for non-members of $\chi$ Persei.}
\begin{tabular}{crcrccccccccc}
\hline
Number&$V$&$b-y$&$m_{1}$&$c_{1}$&$\beta$&$\sigma_{V}$&$\sigma_{b-y}$&$\sigma_{m_{1}}$&$\sigma_{c_{1}}$&$\sigma_{\beta}$&$N_{uvby}$&$N_{\beta}$\\
\hline
2087&13.315& 0.463& 0.002&1.119&2.851&0.003& 0.029&0.045&0.048&0.013&3&3\\
2093&14.863& 0.566&-0.060&0.951&2.831&0.025& 0.012&0.038&0.022&0.036&3&2\\
2097&12.943&0.339&-0.058&0.476&2.699&0.018&0.020&0.028&0.030&0.006&
6&4\\
2098&14.221&0.440&-0.040&1.144&2.891&0.012&0.031&0.044&0.051&0.012&4&3\\
2107&13.894& 1.023& 0.171&0.480&2.604&0.011& 0.016&0.026&0.060&0.010&3&2\\
2127&14.961& 0.666& 0.016&0.491&2.612&0.026& 0.020&0.019&0.039&0.004&3&2\\
2158&15.515& 0.586&-0.025&0.870&0.000&  -  &   -  &  -  &  -  &  -  &1&0\\
2188&15.070& 0.525&-0.036&1.010&2.861&0.018& 0.037&0.056&0.029&0.018&3&2\\
2198&13.487& 1.143& 0.451&0.365&2.608&0.001& 0.008&0.056&0.041&0.023&2&4\\
2202&13.300& 0.500& 0.142&0.337&2.629&0.023& 0.025&0.027&0.019&0.013&3&3\\
2216&14.253& 0.967& 0.161&0.435&2.568&0.024& 0.017&0.009&0.043&0.010&2&2\\
2281&15.110& 0.599&-0.004&0.936&2.749&0.029& 0.026&0.045&0.009&0.005&2&2\\
2315&13.754&0.404&-0.064&1.031&2.833&0.013&0.012&0.024&0.022&0.027&5&4\\
2329&12.762& 0.543& 0.111&0.714&2.657&0.018& 0.022&0.046&0.050&0.005&6&4\\
2345&13.092& 0.202& 0.191&0.307&2.759&0.015& 0.029&0.065&0.051&  -  &3&1\\
2356&15.104& 0.651&-0.001&0.568&2.692&0.013& 0.052&0.105&0.055&0.029&3&2\\
2365&13.701& 0.524& 0.072&0.430&2.620&0.033& 0.035&0.063&0.038&0.018&6&4\\
2370&14.042&0.248&0.111&0.775&2.843&0.055&0.063&0.131&0.090&0.027&
6&4\\
2376&13.481& 0.526& 0.080&0.665&2.717&0.023& 0.034&0.081&0.039&0.030&6&4\\
2381&13.003& 0.525& 0.278&0.389&2.584&0.017& 0.022&0.039&0.041&0.014&6&4\\
2397&14.161& 0.625& 0.036&0.502&2.649&0.029& 0.023&0.072&0.032&0.041&5&4\\
7013&15.320& 0.698&-0.005&0.551&2.601&  -  &   -  &  -  &  -  &0.041&1&2\\
7025&13.292& 0.480&-0.011&0.793&2.777&  -  &  -   &  -  &   - &  -  &1&1\\
7029&16.166& 0.484&-0.061&0.686&2.839&  -  &  -   &  -  &  -  &  -  &1&1\\
7035&15.677& 0.874&-0.161&0.598&2.596&  -  &  -   &  -  &  -  &  -  &1&1\\
7044&16.040& 0.809&-0.110&0.440&2.610&  -  &   -  &  -  &  -  &  -  &1&1\\
7071&15.601& 0.608& 0.004&0.751&2.719&0.002& 0.005&0.005&0.064&0.043&2&2\\
7081&15.899&0.564&0.002&1.233&2.812&-&-&-&-&-&1&1\\
7082&12.856&0.357&0.102 &1.063&  -  &  -  & -   &  -  &  -  &  -  &1&0\\
7101&15.912&0.431&0.024&1.188&2.907&0.042&0.011&0.016&0.036&2&1\\
\hline
\end{tabular}
\end{center}
 \label{tab:nomiembroschi}
\end{table*}

\subsection{Distance Modulus Determination}
\label{sec:distance}

We have estimated the distance modulus to $h$ \& $\chi$ Persei by using
both:(a) the $\beta$ index calibration of Balona \&  Shobbrook (1984) and 
(b) by 
fitting the observed $V_{0}$.vs.$(b-y)_{0}$ and $V_{0}$.vs.$c_{0}$
ZAMS to the
mean calibrations of Perry et al. (1987).

\begin{figure}
\begin{picture}(250,250)
\put(0,0){\includegraphics{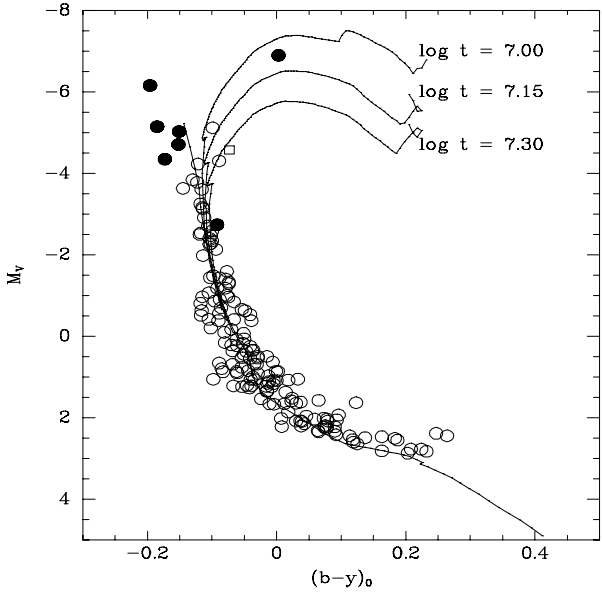}}
\end{picture}
\caption{Absolute magnitude $M_{V}$ against intrinsic colour $(b-y)_{0}$
for $h$ Persei members. The thick line represents the ZAMS from Perry et 
al. (1987). Three isochrones corresponding to $\log t =7.0, 7.15$ and $7.30$
are labelled with their respective $\log t$. Filled circles are supergiant
stars not observed by us and taken from the study of Crawford et al. (1970b). 
Open squares are stars in our sample previously catalogued as having high rotational
velocity (Slettebak 1968).}
\label{fig:869by0}
\end{figure}
\begin{figure}
\begin{picture}(250,250)
\put(0,0){\includegraphics{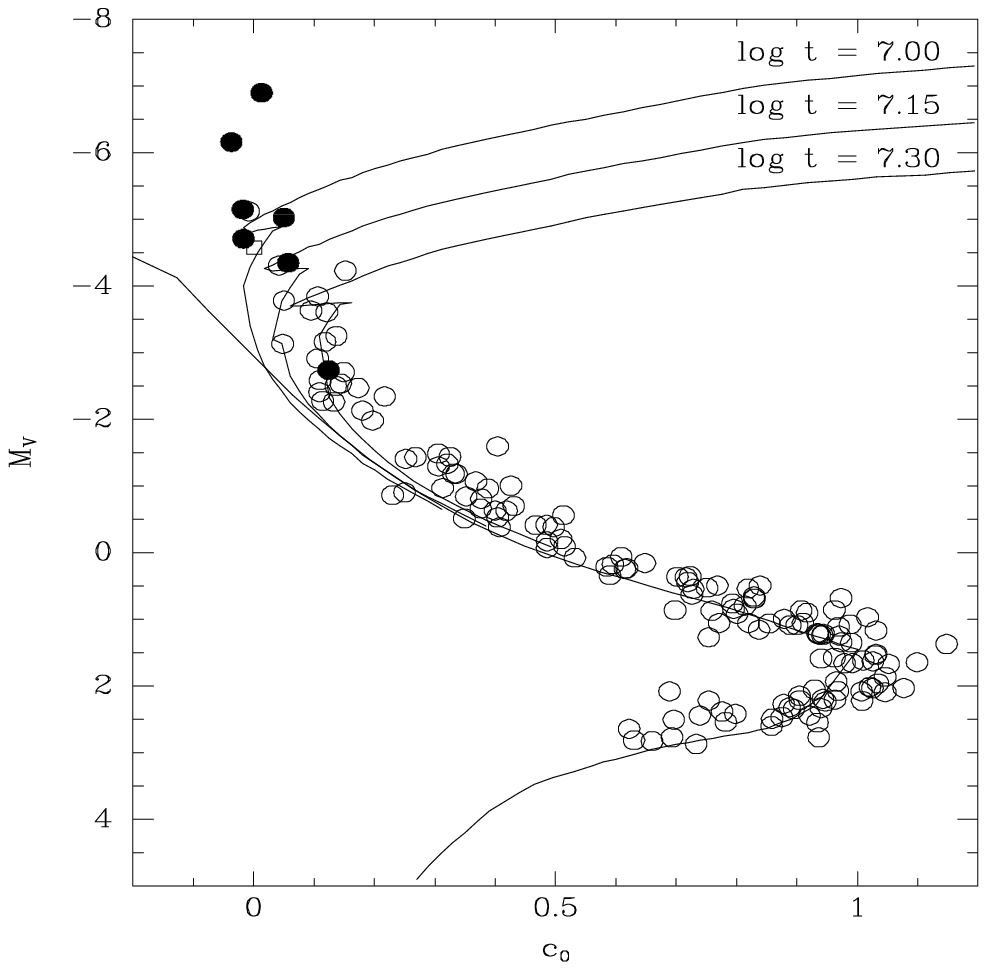}}
\end{picture}
\caption{Absolute magnitude $M_{V}$ against intrinsic colour $c_{0}$
for $h$ Persei members. The thick line represents the ZAMS from Perry et 
al. (1987). Three isochrones corresponding to $\log t =7.0, 7.15$ and $7.30$
are labelled with their respective $\log t$. Filled circles are supergiant
stars not observed by us and taken from the study of Crawford et al. (1970b). 
Open squares are stars in our sample previously catalogued as having high rotational
velocity (Slettebak 1968).}
\label{fig:869c0}
\end{figure}

(a) Since the Balona \& Shobbrook calibration is only valid for
B-type stars,
we select only those stars that we used for the reddening determination,
calculate their intrinsic photometric indices adopting the average reddening
for each cluster, and derive $M_{V}$ from $c_{0}$ and $\beta$. For each star,
we obtain the distance modulus $V_{0} - M_{V}$, and finally 
we
calculate the distance modulus
for each cluster as the average of its  members. The values obtained are
listed in Tables 28 and 29. 

For $h$ Persei, we find an average
$V_{0} - M_{V}= 11.4\pm0.5$, where the uncertainty represents only the
standard deviation of the individual measurements and does not include
the errors derived from the uncertainty in the photometric indices or
the calibration itself. When these are taken into account, the determinations
for all individual stars are compatible with the average value, but for
the possible exception of star 880.

\begin{table*}
\begin{center}
\caption{Photometric data for Be stars of $h$ Persei.}
\begin{tabular}{crcrccccccccc}
\hline
Number&$V$&$b-y$&$m_{1}$&$c_{1}$&$\beta$&$\sigma_{V}$&$\sigma_{b-y}$&$\sigma_{m_{1}}$&$\sigma_{c_{1}}$&$\sigma_{\beta}$&$N_{uvby}$&$N_{\beta}$\\
\hline
922&9.528&0.363&-0.188&0.220&2.600&0.008&0.004&0.015&0.008&  -  &2& 1\\
1161&10.148&0.383&-0.047&0.085&2.559&0.019&0.021&0.046&0.036&0.001&5& 2\\
1268&9.355&0.339&-0.019&0.032&2.624&0.008&0.001&0.008&0.019&  -  &2& 1\\
\hline
\end{tabular}
\end{center}
 \label{tab:miembrosBeh}
\end{table*}

\begin{table*}
\begin{center}
\caption{Photometric data for Be stars of $\chi$ Persei.}
\begin{tabular}{crcrccccccccc}
\hline
Number&$V$&$b-y$&$m_{1}$&$c_{1}$&$\beta$&$\sigma_{V}$&$\sigma_{b-y}$&$\sigma_{m_{1}}$&$\sigma_{c_{1}}$&$\sigma_{\beta}$&$N_{uvby}$&$N_{\beta}$\\
\hline
  2088& 9.117&0.367&-0.161&0.071&2.544&  -  &  -  &   - &   - &   - &1&1\\
  2165&10.053&0.342&-0.114&0.075&2.483&0.018&0.019&0.033&0.016&0.011&4&3\\
  2242&10.956&0.321&-0.076&0.300&2.552&0.010&0.016&0.030&0.021&0.015&5&3\\
  2262&10.538&0.343&-0.110&0.126&2.542&0.005&0.012&0.012&0.014&0.020&5&3\\
  2284& 9.673&0.330&-0.034&-0.074&2.426&0.013&0.017&0.070&0.072&0.019&3&2\\
  2371& 9.204&0.260& 0.040&0.370&2.585&0.022&0.053&0.135&0.513&0.016&3&2\\
\hline
\end{tabular}
\end{center}
 \label{tab:miembrosBechi}
\end{table*}

The average value for $\chi$ Persei is $V_{0} - M_{V}= 12.1\pm0.2$. There 
are several stars that deviate considerably from the cluster mean, namely 
2296, 2091 and 2251, some of which have very large uncertainties in
$M_{V}$. Removing these stars does not significantly change
the average value for $V_{0} - M_{V}$.

\begin{figure}
\begin{picture}(250,250)
\put(0,0){\includegraphics{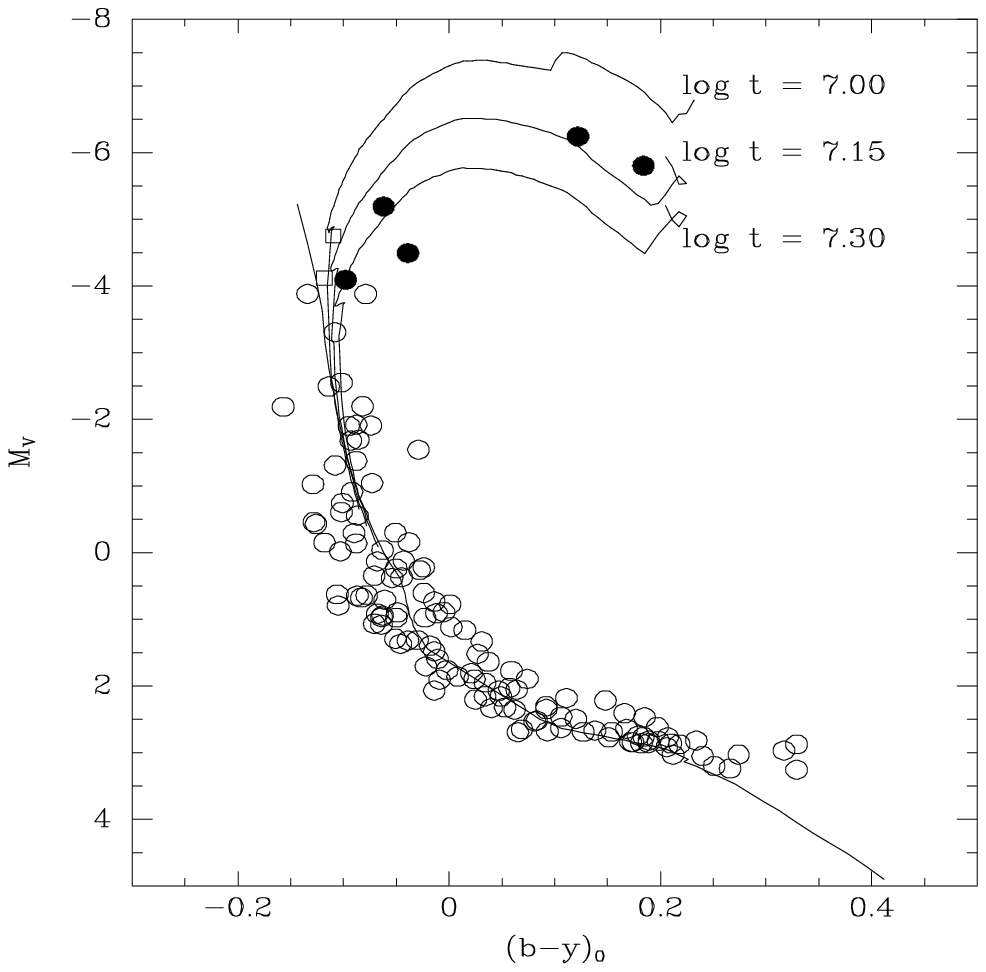}}
\end{picture}
\caption{Absolute magnitude $M_{V}$ against intrinsic colour $(b-y)_{0}$
for $\chi$ Persei members. The thick line represents the ZAMS from Perry et 
al. (1987). Three isochrones corresponding to $\log t =7.0, 7.15$ and $7.30$
are labelled with their respective $\log t$. Filled circles are supergiant and giant
stars not observed by us and taken from the study of Crawford et al. (1970b). Open squares are stars in our sample previously catalogued as having high rotational
velocity (Slettebak 1968).}
\label{fig:884by0}
\end{figure}
\begin{figure}
\begin{picture}(250,250)
\put(0,0){\includegraphics{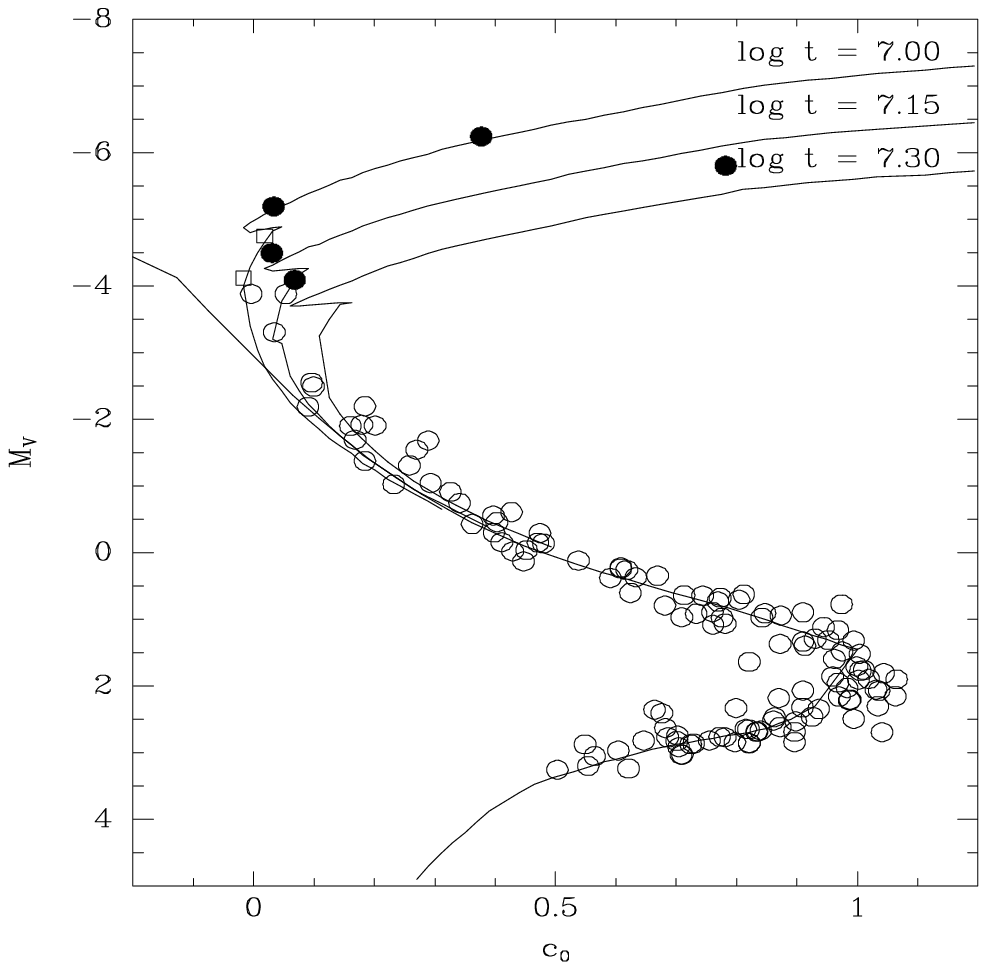}}
\end{picture}
\caption{Absolute magnitude $M_{V}$ against intrinsic colour $c_{0}$
for $\chi$ Persei members. The thick line represents the ZAMS from Perry et 
al. (1987). Three isochrones corresponding to $\log t =7.0, 7.15$ and $7.30$
are labelled with their respective $\log t$. Filled circles are supergiant and giant
stars not observed by us and taken from the study of Crawford et al. (1970b). 
Open squares are stars in our sample previously catalogued as having high rotational
velocity (Slettebak 1968).}
\label{fig:884c0}
\end{figure}

(b) Fitting the theoretical ZAMS is not only intrinsically more accurate, but
also makes use of all the stars in the sample. This is important, because 
when the A and F stars are well fit to the ZAMS, it can be seen
(Figures 9, 10, 11 and 12) that {\em all}
early B stars deviate significantly from the ZAMS. 

We fit individually data from both clusters and derive best fit distance
moduli of $V_{0} - M_{V}= 11.56\pm0.20$ for $\chi$ Persei and 
$V_{0} - M_{V}= 11.66\pm0.20$ for $h$ Persei (the error indicates the uncertainty in 
positioning the theoretical ZAMS and its identification as a lower
envelope). We see that the two values are, within the errors, identical, and
that the values derived by using the Balona \& Shobbrook calibration,
with their larger errors, are compatible with them. We therefore come
to the conclusion that both clusters are indeed at the same distance.

As a further test, we plot in Figures 13 and 14
the combined $V_{0}$\,--\,$(b-y)_{0}$ and  $V_{0}$\,--\,$c_{0}$ diagrams
for stars in both clusters. Points in these diagrams represent average
values for the photometric index, taken over an interval of 0.5 mag,
displaced by the individual distance modulus found for each cluster.
It is obvious that both clusters fit the same ZAMS, confirming that
the distance modulus are basically identical. 

\subsection{Age determination}

(a) We first derive the age of both clusters on the basis of the data of our
sample.
  It is well known that a
  number of giant stars in $h$ \& $\chi$ Persei have rotational
  velocities considerably higher than the average for field stars of
  the same spectral type and also that some of these are Be stars
  (Slettebak 1968; Waelkens et al. 1990; Denoyelle et al. 1994). Such
  stars can occupy positions in the photometric diagrams that differ
  considerably from those of non-peculiar stars of the same spectral
  type. For this reason, no Be stars have been included in our
  plots. Similarly, three supergiants binaries and four possible
  binary candidates in the clusters observed by Abt \& Levy (1973)
  have also been excluded.

\begin{figure}
\begin{picture}(250,250)
\put(0,0){\includegraphics{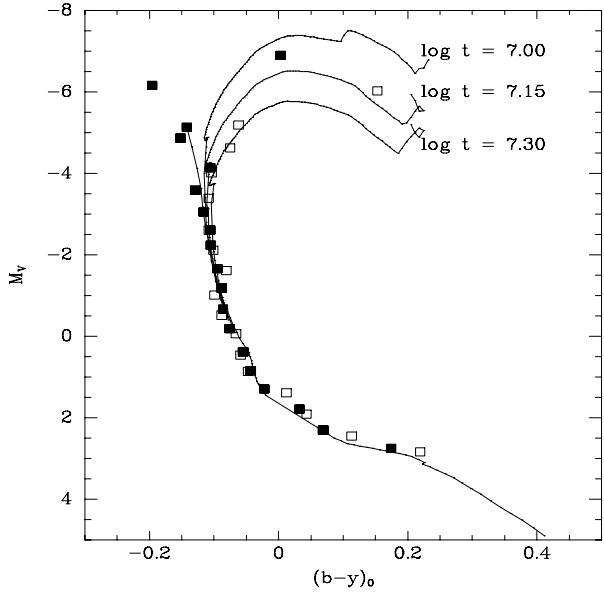}}
\end{picture}
\caption{Absolute magnitude $M_{V}$ against intrinsic colour $(b-y)_{0}$
for $h$ \& $\chi$ Persei members. The ZAMS and isochrones are as in 
figures for the individual clusters. Data points represent averages taken
over 0.5 mag intervals for each cluster. Filled squares are datapoints for
$h$ Persei, while open squares are datapoints for $\chi$ Persei.}
\label{fig:binby0}
\end{figure}

\begin{figure}
\begin{picture}(250,250)
\put(0,0){\includegraphics{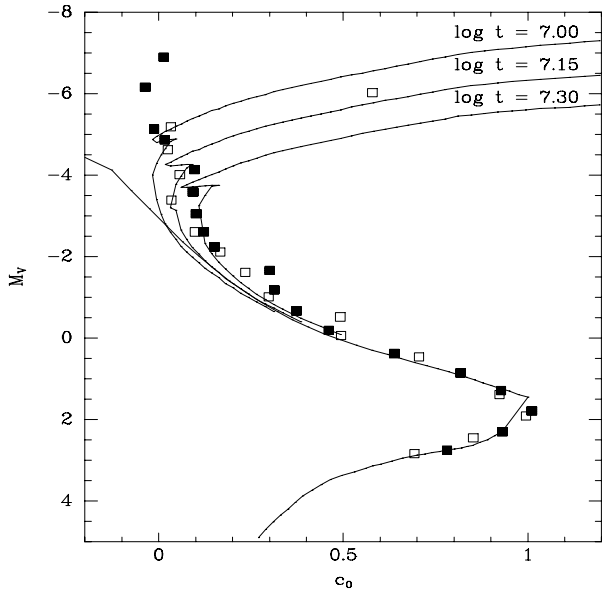}}
\end{picture}
\caption{Absolute magnitude $M_{V}$ against intrinsic colour $c_{0}$
for $h$ \& $\chi$ Persei members. The ZAMS and isochrones are as in 
figures for the individual clusters. Data points represent averages taken
over 0.5 mag intervals for each cluster. Filled squares are datapoints for
$h$ Persei, while open squares are datapoints for $\chi$ Persei.}
\label{fig:binc0}
\end{figure}

By comparing Figures~\ref{fig:binby0} and \ref{fig:binc0}, it is
obvious that the dispersion in the location of evolved stars in the 
$M_{V}-(b-y)_{0}$ diagram is an artifact introduced by differential
reddening, since it disappears completely in the $M_{V}-c_{0}$
diagram. This again confirms that the latter must be preferred
as the reference colour-magnitude plot.

Also plotted in Figures 9, 10, 11, 12, 13 and 14 are isochrones computed 
with the evolutionary
models of Schaller et al. (1992) for a log age of 7.00, 7.15 and 7.30 respectively (Meynet et al. 1993). The metallicities adopted are X=0.68,
Y=0.30 and Z=0.020 (Schaller et al. 1992). The isochrones have been transformed from
the LogT-LogL plane to the $uvby$ system after Torrej\'on (1996). The data
clearly show that the main-sequence turnoff in $h$ Persei occurs at
a later spectral type than in $\chi$ Persei, which indicates that, contrary
to previous speculations, $h$ Persei is actually {\em older} than 
$\chi$ Persei.

In spite of this it is clear from Figure 10 that the brightest stars
in $h$ Persei are substantially younger than the rest of the cluster, with
some of them falling on the $\log t=7.0$ isochrone. This effect cannot be due
to the fact that some of the stars are fast rotators. In Figures 9, 10, 11 and
12 stars with high rotational velocities have been identified (1133 in $h$
Persei and 2296 and 2299 in $\chi$ Persei, Slettebak 1968). Most of the 
stars falling on the $\log t=7.0$ isochrone have low measured rotational
velocities. Morever Meynet \& Maeder (2000) have shown that a $\log t=7.3$ 
isochrone for high rotational velocity stars is almost identical to a 
$\log t=7.2$ isochrone without rotation and therefore a variation in $\log t$ 
of 0.3 dex is too high to be explained by high rotation alone.

(b) In order to verify our conclusions, we have included data for a few stars brighter than $V=11$
  taken from the literature, (see discussion in Section $4.1$). For
  consistency, their unreddened photometric values have been
  calculated with the average reddening obtained for the central region
  of both clusters, since we have already shown that this has little
  bearing on the $M_{V}-c_{0}$ diagram. 

Inclusion of brighter stars in the H-R diagram of $h$ Persei 
corroborates our original finding. All the earliest stars are consistent with
$\log t=7.0$ or even younger, with the brightest supergiant stars even
suggesting $\log t=6.8$ (See Figure 10).

This two-branch age distribution, with most stars being compatible
with $\log t=7.3$ and all the massive stars indicating $\log t=7.0$ or younger
is clearly not due to the presence of binaries or high rotational velocity 
stars.

The presence of stars with high rotation results in an apparent age
dispersion (Meynet \& Maeder 2000) and not in the separation into two branches.
 We note that massive stars from both the central region and the outskirts
of the cluster display this effect without any obvious age separation, i.e., the younger age of massive stars does not seem to be related to their spatial
distribution.

A similar situation can be observed in $\chi$ Persei (see Figure 12).
Two of the brightest stars seem to be younger than the rest of the cluster
and fit the $\log t=7.0$ isochrone. In this case, however, dispersion due
to effects such as rotation cannot be ruled out, specially since not all
bright stars seem younger than the bulk of the cluster.

\section{Discussion}

We find that the reddening and distance moduli to $h$ \& $\chi$ Persei are 
consistent with both clusters being placed at the same distance. However,
the later main-sequence turnoff of $h$ Persei indicates that this cluster
is older than $\chi$ Persei, as far as a single age determination is 
meaningful. From isochrone fitting, we find that the bulk of stars in
$h$ Persei fit an age of $\log t = 7.30$, while in $\chi$
Persei no star seems to be old enough to lie on the $\log t = 7.30$
isochrone. However, all the earliest stars in $h$ Persei deviate
clearly and strongly from the rest of the cluster with some stars falling
along the $\log t=7.0$ isochrone and the brightest objects being even younger
(probably as young as $\approx$ $log t=6.8$.

Almost all the stars in $\chi$ Persei are consistent with 
$\log t=7.10-7.15$, though two of the brightest stars could be slightly
younger ($\log t=7.00$). The low age of the few brightest members of $h$
Persei is the reason
why previous authors attributed a younger age to $h$ Persei than to 
$\chi$ Persei (Tapia  et al. 1984; Schild 1967). On the other hand,
the age of most stars in $\chi$ Persei corresponds approximately to the average between
the two branches in $h$ Persei, which explains why other authors have
given the same age for both clusters (Crawford et al. 1970b). The presence
of at least two branches in the H-R diagram for $h$ Persei strongly suggests
two star formation epochs, the younger one corresponding to the more
massive stars.

Since the age of the bulk of $\chi$ Persei does {\em not} correspond to any of the two isochrone fits
in $h$ Persei (which we find to be at the same distance), the evidence 
points to several stages of star formation in the region.

This effect can be observed both when we consider only the
stars covered by our observations (which are all relatively close
to the main sequence) and also when
the brighter members taken from the literature (in a later evolutionary stage and
not necessarily belonging to the central region) are included. Since we have
excluded any star that could be suspect of binarity or any
pecularities, and the $M_{V}-c_{0}$ is not significatively affected by
reddening, we may conclude that the age spread is real.

Our distance determination is consistent with some of the higher values
found in literature (except those which give a different and larger
distance to $\chi$ Persei). We derive our distance by fitting the ZAMS to
stars much fainter than in previous work. As indicated by Vrancken et al.
(2000), the lower distance moduli measured by Crawford et al. (1970b)
and Balona \& Shobbrook (1984) are due to their use of only the brightest
stars. As can be seen in our HR diagrams, all stars earlier than
$\approx$ B3 deviate considerably from the ZAMS. This is again in
agreement with the results of Vrancken et al. (2000), who find that all
the stars in their sample of B1 and B2 stars are giants, even though
some of them were previously classified as main-sequence.

From our data, we find no new Be stars in $h$ \& $\chi$ Persei. This is
not the last word on this issue, because many catalogued Be stars do 
actually show
a $\beta$ index that does not indicate emission in our data. Since our
census of B stars in the areas observed is complete, we can 
calculate the fraction of Be stars with respect to
total number of B stars. In 
$h$ Per, we find 3 Be stars among 74 B stars, which means an abundance
$N_{\rm Be}/N_{{\rm B + Be}}$ (supergiants excluded) of 4\%. In $\chi$ Persei, 
we find 6 Be stars out of 53 B stars, representing an abundance of 11\%.
Given the scatter in ages in $h$ Per and the small number of Be stars,
we cannot derive any conclusions about the effect of cluster age on
Be abundance.


\section{Conclusions}
We derive reddening and distance values consistent with the idea that
$h$ Persei and $\chi$ Persei are placed at the same distance. From the
ZAMS fitting, we derive an approximate distance modulus 
$V_{0}-M_{V}=11.6\pm0.2$ for both clusters. The ages of the two clusters
seem to be, however, different. There is evidence for two massive star
populations in $h$ Persei, fitting the $\log t = 7.0$ and $\log t=7.3$
isochrones, with the more massive stars being younger. The age spread in
$\chi$ Persei is, on the other hand, negligible, with all stars fitting
the $\log t =7.10-7.15$ isochrone. We interpret these data as favouring the
idea that both clusters belong to a single star forming region in which
at least three different star formation stages have taken place.

\section*{Acknowledgements}

We would like to thank the Spanish CAT panel for allocating observing time
to this project. AM would like to  thank Dr.~J.~Fabregat for his help with the
observations and  Dr.~J.~M.~Torrej\'{o}n for making
available the theoretical isochrones transformed to the $M_{V}/c_{0}$ and 
$M_{V}/(b-y)_{0}$ spaces.

\begin{table*}[!]
\begin{center}
\caption{Intrinsic photometric values and approximate spectral classification
for all members ($h$ Persei members on the left column, $\chi$ Persei members
on the right). Unless otherwise indicated in the text, luminosity class is always taken to be V.}
\begin{tabular}{crrrc|crrrc}
\hline
Number&$V_{0}$&$(b-y)_{0}$&$c_{0}$&Spectral Type&Number&$V_{0}$&$(b-y)_{0}$&$c_{0}$&
Spectral Type\\
\hline
 820&11.281&-0.090&0.497&B6&2085& 9.648&-0.088&0.179&B3\\
 821&13.600&0.096&0.965&A8&2091&10.022&-0.029&0.270&B3\\
 832&12.325&-0.089&0.829&A0&2092&13.086&0.027&1.004&A2\\
 836&13.883&0.008&0.754&A3&2094&10.189&-0.088&0.184&B2\\
 837&12.203&-0.029&0.819&A0&-&-&-&-&-\\
 842&11.250&-0.105&0.485&B7&2108&12.936&-0.046&0.872&A0\\
 843& 7.438&-0.122&0.152&B1&2109&13.340&0.059&1.005&A5\\
 844&13.700&0.059&1.025&A8&2111&11.787&-0.024&0.608&B7\\
 845&12.872&-0.046&0.934&A0&2114& 9.369&-0.082&0.184&B2\\
 848&13.203&-0.024&1.030&A2&2116&11.826&-0.028&0.618&B8\\
 854&12.348&-0.035&0.973&A1&2123&13.376&0.021&1.044&A2\\
 856&13.699&0.078&1.077&A7&2124&13.462&0.074&1.019&A5\\
 857&13.017&-0.015&0.990&A0&2133&10.521&-0.073&0.293&B3\\
 864& 8.035&-0.145&0.095&B2&2139& 9.661&-0.074&0.201&B1\\
 875&13.719&0.093&0.929&A5&2147&12.682&0.002&0.944&A0\\
 876&10.823&-0.066&0.352&B3&2149&12.456&-0.005&0.760&B9\\
 879& 9.686&-0.114&0.197&B2&2167&11.687&-0.043&0.538&B6\\
 880&11.137&-0.041&0.404&B3&2170&13.621&0.064&1.031&A5\\
 885&13.870&0.038&0.963&A5&2174&13.638&0.047&1.036&A7\\
 892& 9.324&-0.100&0.217&B1&2175&12.878&-0.030&0.952&A0\\
 893&10.072&-0.077&0.404&B4&2179&12.462&-0.049&0.910&A0\\
 896&10.597&-0.105&0.368&B5&2193&13.725&0.049&0.970&A5\\
 898&12.660&-0.037&0.879&A0&2194&11.804&-0.050&0.608&B8\\
 901&13.010&-0.015&0.973&A0&2196& 9.872&-0.086&0.168&B2.5\\
 907&10.482&-0.092&0.331&B3&2200&11.011&-0.087&0.397&B5\\
 909&13.296&0.123&1.026&A7&2206&13.913&0.092&0.936&A7\\
 914&13.310&0.030&1.099&A3&2209&13.719&0.034&1.063&A5\\
 917&12.883&-0.067&0.935&A1&2211&11.273&-0.090&0.474&B7\\
 923&11.152&-0.117&0.349&B5&2214&14.097&0.081&0.898&A7\\
 924&13.281&0.037&1.010&A3&2215&13.869&0.092&1.034&A5\\
 929& 8.414&-0.118&0.137&B2&2219&13.746&0.111&0.870&A7\\
 930&10.485&-0.079&0.336&B5&2223&12.341&0.001&0.974&A1\\
 934&12.753&0.000&0.889&A0&2224&12.169&-0.024&0.624&B8\\
 935&12.159&-0.029&0.768&B9&2229& 9.666&-0.095&0.160&B2\\
 936& 8.503&-0.115&0.118&B1&2232& 9.377&-0.157&0.090&B2\\
 939&10.371&-0.074&0.306&B3&2235& 7.688&-0.079&0.053&B1\\
 941&13.325&-0.011&0.979&A1&2239&12.543&-0.023&0.776&B9\\
 945&12.904&-0.053&0.942&A1&2240&12.359&-0.105&0.681&A0\\
 946&12.526&-0.001&0.962&A1&2241&11.913&-0.071&0.669&B9\\
\hline
\end{tabular}
\end{center}
 \label{tab:intrinsic1}
\end{table*}
\begin{table*}[!]
\begin{center}
\caption{Intrinsic photometric values and approximate spectral classification
for all members ($h$ Persei members on the left column, $\chi$ Persei members
on the right). Unless otherwise indicated in the text, luminosity class is always taken to be V. (Continued Table 26).}
\begin{tabular}{crrrc|crrrc}
\hline
Number&$V_{0}$&$(b-y)_{0}$&$c_{0}$&Spectral Type&Number&$V_{0}$&$(b-y)_{0}$&$c_{0}$&
Spectral Type\\
\hline
 947&13.936&0.077&0.878&A7&2245&10.820&-0.101&0.341&B5\\
 948&13.741&0.037&0.968&A3&2246& 8.259&-0.108&0.034&B1\\
 949&13.812&0.043&0.904&A5&2249&13.472&-0.009&1.002&A2\\
 950& 9.405&-0.104&0.133&B2&2251& 9.886&-0.093&0.289&B4\\
 952&10.180&-0.098&0.306&B3&2253&10.956&-0.102&0.427&B7\\
 955&13.889&0.090&1.008&A8&2254&13.595&0.057&0.983&A5\\
 956&10.662&-0.077&0.426&B5&2255& 9.015&-0.102&0.096&B2\\
 959&10.966&-0.086&0.431&B6&2258&12.271&-0.061&0.804&B9\\
 960&11.823&-0.080&0.648&B9&2260&12.300&-0.014&0.769&B9\\
 963& 9.129&-0.117&0.144&B1&2261&12.857&-0.051&0.930&A0\\
 965&10.703&-0.075&0.388&B5&2267&11.547&-0.103&0.429&B7\\
 966&12.354&-0.068&0.830&A0&2268&11.945&-0.054&0.591&B8\\
 970&12.511&-0.044&0.795&B9&2269&11.427&-0.088&0.480&B7\\
 971&12.568&-0.061&0.917&A0&2270&12.476&-0.012&0.847&A0\\
 978& 8.754&-0.112&0.106&B2&2275&11.409&-0.038&0.411&B6\\
 979&12.354&-0.031&0.829&A0&2277&13.271&-0.022&0.999&A2\\
 980& 7.824&-0.130&0.106&B1.5&2283&12.963&-0.018&0.913&A0\\
 982&12.027&-0.068&0.702&B9&2286&13.048&-0.014&0.975&A1\\
 985&10.226&-0.103&0.325&B4&2294&12.481&-0.068&0.733&B9\\
 986&10.697&-0.115&0.313&B5&2296& 6.816&-0.110&0.018&B1.5\\
 987&12.743&0.018&0.988&A3&2297&11.266&-0.051&0.398&B5\\
 988&10.857&-0.118&0.377&B6&2299& 7.447&-0.118&-0.01&B1\\
 990&12.537&-0.083&0.759&A0&2300&13.921&0.062&0.664&-\\
 991& 9.535&-0.094&0.180&B2.5&2301&10.257&-0.108&0.258&B5\\
 992& 8.060&-0.116&0.121&B1.5&2307&13.518&0.034&0.968&A5\\
 997& 9.194&-0.102&0.173&B2&2309&11.107&-0.128&0.403&B7\\
 999&11.462&-0.102&0.509&B8&2311& 7.713&-0.106&0.152&B2\\
1004& 8.957&-0.106&0.149&B2&2314&12.883&-0.039&0.994&A1\\
1007&13.683&0.007&1.021&A5&2317&13.328&-0.002&1.011&A2\\
1014&11.034&-0.116&0.400&B7&2319&11.415&-0.118&0.471&B8\\
1017&13.899&0.074&0.947&A8&2323&12.541&-0.050&0.842&A0\\
1018&13.761&0.039&1.046&A5&2324&12.212&-0.087&0.744&A0\\
1020&11.879&-0.070&0.585&B8&2331&12.508&-0.063&0.873&A0\\
1021&11.105&-0.089&0.513&B8&2332&12.901&0.031&0.910&A2\\
1025&13.859&0.075&0.943&A5&2335&12.729&0.015&0.968&A2\\
1028&12.886&-0.039&0.944&A1&2338&11.933&-0.045&0.633&B9\\
1030&14.156&0.137&0.859&A7&2349&11.138&-0.126&0.362&B7\\
1031&13.324&0.012&0.992&A3&2350&11.699&-0.068&0.447&B7\\
1034&13.254&0.023&0.940&A3&2352&10.655&-0.092&0.326&B6\\
\hline
\end{tabular}
\end{center}
 \label{tab:intrinsic1}
\end{table*}

\begin{table*}[!]
\begin{center}
\caption{Intrinsic photometric values and approximate spectral classification
for all members ($h$ Persei members on the left column, $\chi$ Persei members
on the right). Unless otherwise indicated in the text, luminosity class is always taken to be V. (Continued Table 26).}
\begin{tabular}{crrrc|crrrc}
\hline
Number&$V_{0}$&$(b-y)_{0}$&$c_{0}$&Spectral Type&Number&$V_{0}$&$(b-y)_{0}$&$c_{0}$&Spectral Type\\
\hline
1038&13.678&0.073&1.021&A5&2355&13.897&0.040&0.799&A5\\
1041& 9.165&-0.119&0.136&B2.5&2358&12.631&-0.071&0.781&A0\\
1049&12.005&-0.036&0.590&B7&2359&12.205&-0.078&0.713&B9\\
1050&14.110&0.113&0.921&A7&2362&13.769&0.025&0.988&A3\\
1052&13.334&-0.004&1.052&A2&2363&12.243&-0.083&0.773&A0\\
1053&12.721&-0.098&0.771&A0&-&-&-&-&-\\
1056&12.779&-0.012&0.969&A0&2379&10.541&-0.129&0.232&B5\\
1058&11.729&-0.050&0.609&B8&2392& 9.075&-0.114&0.099&B2\\
1059&12.836&-0.005&1.031&A1&2401&13.781&0.148&0.987&A5\\
1064&12.721&-0.040&0.820&A0&2407&12.647&-0.064&0.761&A0\\
1066&11.253&-0.066&0.467&B7&2410&12.533&-0.064&0.709&A0\\
1077&11.900&-0.061&0.618&B9&2414&12.192&-0.106&0.812&A2\\
1078& 7.883&-0.123&0.050&B2&2416&13.636&-0.014&0.910&A3\\
1079&13.747&0.030&0.689&A2&7014&14.210&0.168&0.815&F0\\
1080& 9.259&-0.104&0.109&B2.5&7015&14.253&0.154&0.833&A7\\
1081&12.428&-0.054&0.793&A0&7016&14.026&0.106&0.925&A5\\
1083&11.496&-0.070&0.486&B7&7017&14.486&0.206&0.7047&F3\\
1085& 8.534&-0.114&0.048&B1.5&7021&14.594&0.274&0.708&F4\\
1091&13.986&0.064&0.888&A7&7023&14.383&0.234&0.755&A8\\
1093&11.594&-0.051&0.486&B7&7024&14.329&0.184&0.772&A8\\
1095&11.743&-0.062&0.532&B8&7027&14.036&0.185&0.863&F0\\
1096&13.186&0.024&1.031&A3&-&-&-&-&-\\
1105&12.107&-0.051&0.719&B9&7037&14.618&0.240&0.565&-\\
1106&12.212&-0.041&0.728&B9&7038&13.466&0.024&1.065&A2\\
1108&12.037&-0.047&0.716&B9&7045&13.975&0.166&0.676&F0\\
1109& 9.080&-0.100&0.110&B3&7046&14.260&0.065&1.041&A8\\
1110&11.569&-0.081&0.515&B8&7047&14.425&0.182&0.821&A7\\
1116& 7.360&-0.089&0.042&B2&7048&14.058&0.120&0.994&A5\\
1117&13.877&0.078&0.905&A8&7049&14.804&0.266&0.621&-\\
1118&12.193&-0.054&0.751&B9&7052&14.386&0.189&0.646&A7\\
1121&11.841&-0.052&0.595&B8&7054&14.392&0.198&0.701&F2\\
1122&10.332&-0.074&0.321&B6&7062&13.425&0.008&0.959&A2\\
1126&10.803&-0.096&0.230&B5&7064&14.255&0.127&0.896&F0\\
1128&10.261&-0.079&0.252&B5&7066&14.248&0.093&0.834&-\\
1129&11.038&-0.049&0.419&B7&7067&14.597&0.212&0.710&F5\\
1130&12.582&-0.029&0.801&A0&7068&14.233&0.138&0.840&A8\\
1132& 6.548&-0.099&-0.00&B2&7070&14.441&0.218&0.724&F0\\
1133& 7.095&-0.073&0.001&B1&7077&14.421&0.189&0.821&F0\\
1145&12.894&-0.016&0.937&A2&7080&14.182&0.197&0.872&A7\\
\hline
\end{tabular}
\end{center}
 \label{tab:intrinsic2}
\end{table*}

\begin{table*}[!]
\begin{center}
\caption{Intrinsic photometric values and approximate spectral classification
for all members ($h$ Persei members on the left column, $\chi$ Persei members
on the right). Unless otherwise indicated in the text, luminosity class is always taken to be V. (Continued Table 26).}
\begin{tabular}{crrrc|crrrc}
\hline
Number&$V_{0}$&$(b-y)_{0}$&$c_{0}$&Spectral Type&Number&$V_{0}$&$(b-y)_{0}$&$c_{0}$&Spectral Type\\
\hline
1147&12.719&0.033&0.910&A3&7083&13.889&0.053&0.909&A7\\
1152&12.909&-0.008&0.970&A1&7084&11.529&-0.063&0.452&B7\\
1163&13.240&0.065&0.962&A7&7085&14.197&0.106&0.682&A7\\
1175&12.636&-0.013&1.017&A1&7086& 7.684&-0.134&-0.00&B2.5\\
1179&11.003&-0.054&0.376&B6&7088&14.823&0.329&0.503&-\\
1180&12.937&-0.042&0.754&A2&7091&14.535&0.317&0.604&F2\\
1181&10.763&-0.088&0.250&B5&7092&13.162&-0.011&0.962&A2\\
1185&11.286&-0.039&0.407&B7&7093&14.337&0.151&0.686&F0\\
1190&13.527&0.018&1.046&A3&7096&14.310&0.179&0.702&A8\\
1191&12.821&-0.014&0.837&A0&7097&14.442&0.329&0.549&F4\\
1192&13.037&0.014&1.148&A3&7099&14.428&0.210&0.729&F2\\
1198&11.912&-0.041&0.615&B9&7104&14.086&0.083&0.861&A7\\
1202&10.231&-0.088&0.268&B5&7105&14.406&0.172&0.896&F0\\
1203&12.018&-0.038&0.723&B9&7108&14.766&0.251&0.554&F2\\
1206&12.754&-0.005&0.900&A1&7109&14.408&0.174&0.797&F0\\
1213&12.730&-0.034&0.854&A0&7116&14.334&0.207&0.781&F0\\
1218&13.997&0.090&0.940&A7&7118&14.216&0.069&0.820&A8\\
1222&13.747&0.079&1.007&A5&7122&13.204&0.037&0.821&A3\\
1232& 9.391&-0.102&0.114&B3&   &      &     &     \\
1240&14.013&0.065&0.895&A7&   &      &     &     \\
1251&13.626&0.046&1.034&  -  &      &     &     \\
1260&12.296&-0.006&0.726&A0&    &      &     &     \\
1262&12.531&-0.062&0.907&A0&    &      &     &     \\
1265&12.462&-0.085&0.814&A0&    &      &     &     \\
1267&12.530&0.002&0.698&A0&    &      &     &     \\
1281&12.159&-0.015&0.839&A0&    &      &     &     \\
4009&14.050&0.247&0.776&A8&    &      &     &     \\
4011&14.084&0.092&0.798&A7&    &      &     &     \\
4012&14.108&0.264&0.739&F2&    &      &     &     \\
4013&14.131&0.163&0.875&A8&    &      &     &     \\
4016&14.171&0.183&0.696&F2&    &      &     &     \\
4017&14.205&0.118&0.782&F0&    &      &     &     \\
4018&14.216&0.187&0.934&A8&    &      &     &     \\
4023&14.266&0.119&0.858&F0&    &      &     &     \\
4025&14.309&0.125&0.622&F0&    &      &     &     \\
4029&14.438&0.224&0.936&A7&    &      &     &     \\
4030&14.438&0.208&0.693&  -  &      &     &     \\
4036&14.480&0.163&0.630&F0&    &      &     &     \\
4037&14.493&0.232&0.660&F0&    &      &     &     \\
4042&14.538&0.203&0.733&F0&    &      &     &     \\
\hline
\end{tabular}
\end{center}
 \label{tab:intrinsic3}
\end{table*}

\begin{table*}[!]
\begin{center}
\caption{Individual intrinsic data for B stars in $h$ (right) and $\chi$
(left) Persei. The colour excess $E(b-y)$ has been calculated using Crawford's et al.
(1970b) procedure. The absolute magnitudes have been calculated by using
the calibration (based on the $\beta$ index) by Balona \& Shobbrook (1984).
The error in $M_{V}$ has been computed using the formula by Balona \& Shobbrook
(1984). See text for details.}
\begin{tabular}{crrrr|crrrr}
\hline
Number&$E(b-y)$&$M_{V}$&$\sigma_{M_{V}}$&$V_{0}-M_{V}$&Number&$E(b-y)$&$M_{V}$&$\sigma_{M_{V}}$&$V_{0}-M_{V}$\\
\hline
 2085&0.401&-1.747&0.509&11.333& 843& 0.419&  -   & -   &  -    \\
 2091&0.452&-4.392&0.471&14.147& 864& 0.401&-2.224&0.298&10.259\\
 2094&0.400&-1.858&0.252&12.003& 876& 0.456&-0.953&0.140&11.776\\
 2111&0.424&-0.419&0.029&12.061& 879& 0.423&-2.051&0.273&11.737\\
 2114&0.406&-2.223&0.131&11.521& 880& 0.477&-1.539&0.169&12.676\\
 2116&0.419&-0.334&0.230&12.037& 892& 0.435&-2.015&0.542&11.339\\
 2133&0.405&-1.401&0.263&11.858& 893& 0.440&-1.037&0.254&11.109\\
 2139&0.413&-2.326&0.164&11.888& 896& 0.415&-1.384&0.405&11.981\\
 2167&0.411&-0.377&0.103&11.972& 907& 0.432&-1.222&0.345&11.704\\
 2194&0.397&-0.391&0.438&12.164& 929& 0.424&-2.647&0.120&11.061\\
 2196&0.404&-2.135&0.252&11.947& 930& 0.444&-0.888&0.120&11.373\\
 2200&0.380&-0.841&0.136&11.893& 936& 0.429&-2.673&0.090&11.176\\
 2211&0.370&-0.916&0.365&12.276& 939& 0.453&-1.221&0.382&11.592\\
 2224&0.422&-0.107&0.335&12.138& 950& 0.439&-2.533&-&11.938\\
 2229&0.396&-2.228&0.312&11.870& 952& 0.428&-1.465&0.125&11.645\\
 2232&0.339&-2.383&0.502&11.978& 956& 0.438&-0.919&0.087&11.581\\
 2235&0.422&-3.839&0.968&11.387& 959& 0.428&-0.954&0.371&11.920\\
 2241&0.370&0.109&0.388&11.892& 963& 0.425&-2.331&0.027&11.460\\
 2245&0.372&-1.520&0.106&12.419& 965& 0.443&-0.844&0.493&11.547\\
 2246&0.395&-3.555&0.904&11.794& 978& 0.434&-2.534&0.085&11.288\\
 2251&0.385&-1.004&0.211&10.912& 980& 0.415&-2.788&0.280&10.612\\
 2253&0.362&-0.496&0.426&11.572& 985& 0.421&-1.156&0.113&11.382\\
 2255&0.395&-3.201&0.287&12.196& 986& 0.410&-1.280&0.078&11.977\\
 2267&0.361&-0.300&0.199&11.972& 988& 0.401&-0.814&0.152&11.671\\
 2268&0.395&-0.382&0.511&12.306& 991& 0.445&-2.224&0.131&11.759\\
 2269&0.371&-0.560&0.280&12.068& 992& 0.428&-2.310&-&10.370\\
 2275&0.429&-0.689&0.599&11.931& 997& 0.437&-2.023&0.146&11.217\\
 2296&0.394&-3.855&0.825&10.652&1004& 0.435&-2.327&0.107&11.284\\
 2297&0.417&-1.003&0.108&12.153&1014& 0.400&-0.442&0.386&11.476\\
 2299&0.390&-3.860&0.292&11.308&1021& 0.417&-0.085&0.254&11.190\\
\hline
\end{tabular}
\end{center}
 \label{tab:balona1}
\end{table*}
\begin{table*}[!]
\begin{center}
\caption{Individual intrinsic data for B stars in $h$ (right) and $\chi$
(left) Persei. The colour excess $E(b-y)$ has been calculated using Crawford's et al.
(1970b) procedure. The absolute magnitudes have been calculated by using
the calibration (based on the $\beta$ index) by Balona \& Shobbrook (1984).
The error in $M_{V}$ has been computed using the formula by Balona \& Shobbrook
(1984). See text for details. (Continued Table 28).}
\begin{tabular}{crrrr|crrrr}
\hline
Number&$E(b-y)$&$M_{V}$&$\sigma_{M_{V}}$&$V_{0}-M_{V}$&Number&$E(b-y)$&$M_{V}$&$\sigma_{M_{V}}$&$V_{0}-M_{V}$\\
\hline
 2301&0.373&-1.185&0.371&11.516&1041& 0.423&-2.199&0.406&11.364\\
 2309&0.338&-0.935&0.123&12.266&1078& 0.428&-3.191&0.035&11.074\\
 2319&0.341&-0.590&0.299&12.214&1080& 0.441&-2.284&0.102&11.543\\
 2324&0.346&-0.095&0.091&12.496&1085& 0.437&-2.839&0.090&11.373\\
 2338&0.400&0.253&0.180&11.638&1109& 0.445&-2.172&0.269&11.252\\
 2349&0.344&-0.896&0.104&12.232&1116& 0.463&-3.056& -   &10.416\\
 2350&0.395&-0.711&0.114&12.389&1122& 0.451&-0.819&0.149&11.151\\
 2352&0.382&-1.013&0.268&11.702&1126& 0.438&-1.146&0.430&11.949\\
 2359&0.358&0.487&0.144&11.855&1128& 0.453&-1.487&0.330&11.748\\
 2379&0.354&-1.367&0.157&12.064&1129& 0.467&-0.642&0.175&11.680\\
 2392&0.382&-2.142&0.383&11.250&1132& 0.458&-4.037& -   &10.585\\
 2414&0.320&0.424&-&12.069&1133& 0.484&  -   & -   &  -    \\
 7084&0.399&  -   &  -   & -    &1179& 0.466&-1.232&0.077&12.235\\
 7086&0.372&-3.240&0.165&11.001&1181& 0.444&-1.216&0.224&11.979\\
 -&-&-&-&-&1202& 0.442&-1.132&0.109&11.363\\
 -&-&-&-&-&1232& 0.443&-2.456&0.110&11.847\\
\hline
\end{tabular}
\end{center}
 \label{tab:balona1}
\end{table*}

\begin{thebibliography}{99}

\bibitem[1973]{abt73} Abt H.A., Levy S.G., 1973, ApJ 184, 167

\bibitem[1973]{allen73} Allen C.W., 1973, Astrophysical Quantities, 3rd ed. (London, Athlone)

\bibitem[1984]{bal84} Balona L.A., Shobbrook R.R., 1984, MNRAS 211, 375

\bibitem[1943]{bidel43} Bidelman W.P., 1943, ApJ 98, 61

\bibitem[1979]{canter79} Canterna R., Perry C.L., Crawford D.L., 1979, PASP 91, 263

\bibitem[1975]{craw75a} Crawford D.L., 1975a, AJ 80, 955

\bibitem[1975]{craw75b} Crawford D.L., 1975b, PASP 87, 481

\bibitem[1978]{craw78} Crawford D.L., 1978, AJ 83, 48

\bibitem[1994]{craw94} Crawford D.L., 1994, PASP 106, 397

\bibitem[1970]{craw70a} Crawford D.L., Barnes J.V., 1970a, AJ 75, 978

\bibitem[1966]{craw66} Crawford D.L., Mander J., 1966, AJ 71, 114

\bibitem[1976]{craw76} Crawford D.L., Mandwewala N., 1976, PASP 88, 917

\bibitem[1977]{craw77} Crawford D.L., Barnes J.V., Hill G., 1977, AJ 82, 606

\bibitem[1970]{craw70b} Crawford D.L., Glaspey J.W., Perry C.L., 1970b, AJ 75, 822

\bibitem[1994]{denoy94} Denoyelle J., Aerts C., Waelkens C., 
1994, In: Balona L., Heinrich H., Lecontel J. M. (eds.) IAU Symp.\ 162, Pulsation, Rotation and Mass Loss in Early-Type Stars. Kluwer Academic Press, Holland, p.151-152

\bibitem[1985]{doom85} Doom C., De Greve J.P., De Loore C., 1985, ApJ 290, 185

\bibitem[1996]{fat96} Fabregat J., Torrej\'{o}n J.M., Reig P. et al., 1996, A\&AS 119, 271 

\bibitem[1976]{gronb76} Gr\o nbech B., Olsen E.H., Str\"{o}mgren B., 1976, A\&AS 26, 155 

\bibitem[1965]{hoag65} Hoag A.A., Applequist L., 1965, ApJS 12, 215

\bibitem[1955]{john55} Johnson H.L., Morgan W.W., 1955, ApJ 122, 429

\bibitem[1980]{KaE80} Kontizas E., Theodossiou E., 1980, MNRAS 192, 745

\bibitem[1997]{Krze99} Krzesi\'{n}ski J., Pigulski A., 1997, A\&A 325, 987

\bibitem[1986]{lester86} Lester J.B., Gray R.O., Kurucz R.L., 1986, ApJS 61, 509

\bibitem[1993]{manf93} Manfroid J., 1993, A\&A 271, 714

\bibitem[1987]{manf87} Manfroid J., Sterken C., 1987, A\&A 71, 539

\bibitem[1967]{mendoza67} Mendoza E.E., 1967, Bol. Obs. Tonantzintla y Tacubaya 4, 149

\bibitem[2000]{meynet00} Meynet G., Maeder A., 2000, A\&A 361, 101

\bibitem[1993]{meynet93} Meynet G., Mermilliod J.C., Maeder A., 1993, A\&AS 98, 477

\bibitem[1985]{moon85} Moon T.T., Dworetsky M.M., 1985, MNRAS 217, 305

\bibitem[1956]{morgan56} Morgan W.W., Harris D.L., 1956, Vistas Astron. 2, 1124

\bibitem[1983]{muminov83} Muminov M., 1983, BICDS 24, 95

\bibitem[1993]{napi93} Napiwotzki R., Sch\"{o}nberner, Wenske V., 1993, A\&A 268, 653

\bibitem[1937]{oost37} Oosterhoff P.T., 1937, Ann. Sterrewatch Leiden 17, 1

\bibitem[1977]{perry78} Perry C.L., Lee P.D., Barnes J.V., 1978, PASP 90, 73

\bibitem[1987]{perry87} Perry C.L., Olsen E.H., Crawford D.L., 1987, PASP 99, 1184

\bibitem[1992]{schaller92} Schaller G., Schaerer D., Meynet G., Maeder A., 1992, A\&A 96, 269

\bibitem[1965]{shi65} Schild R.E., 1965, ApJ 142, 979

\bibitem[1966]{shi66} Schild R.E., 1966, ApJ 146, 142

\bibitem[1967]{shi67} Schild R.E., 1967, ApJ 148, 449

\bibitem[1976]{sar76} Schild R.E., Romanishin W., 1976, ApJ 204, 493

\bibitem[1983]{shob83} Shobbrook R.R., 1983, MNRAS 205, 1215

\bibitem[1968]{slet68} Slettebak A., 1968, ApJ 154, 933

\bibitem[1987]{stet87} Stetson P.B., 1987, PASP 99, 191

\bibitem[1984]{tapia84} Tapia M., Roth M., Costero R., Navarro S., 1984, Rev. Mex. Astron. Astrofis. 9, 65  

\bibitem[1996]{torrejon} Torrej\'on J.M., 1996, Tesina de Licenciatura, University of Valencia

\bibitem[2000]{vran00} Vrancken M., Lennon D.J., Dufton P.L., Lambert D.L., 2000, A\&A 358, 639

\bibitem[1990]{waelk90} Waelkens C., Lampens P., Heynderickx D. et al., 1990, A\&AS 83, 11

\bibitem[1964]{wildey64} Wildey R.L., 1964, ApJS 8, 439

\end{thebibliography}
\end{document}